\documentclass[iop]{emulateapj}
\usepackage{float}
\usepackage{color}
\usepackage{booktabs}
\usepackage{amsmath}
\newcommand{\ra}[1]{\renewcommand{\arraystretch}{#1}}

\newcommand{\Msol}{\hbox{${\rm M_\odot}$}}
\newcommand{\Lsol}{\hbox{${\rm L_\odot}$}}
\newcommand{\uJy}{\hbox{{\rm $\mu$Jy}}}
\newcommand{\um}{\hbox{{\rm $\mu$m}}}

\begin{document} 

\shortauthors{Wang et al.}
\shorttitle{Infrared color selection of massive galaxies at $z \gtrsim 3$}
\submitted{ApJS, accepted on Nov 25th, 2015}
\title{Infrared color selection of massive galaxies at $z \gtrsim 3$}
\author{T.~Wang$\!$\altaffilmark{1},
D.~Elbaz$\!$\altaffilmark{1},
C.~Schreiber$\!$\altaffilmark{1}, 
M.~Pannella$\!$\altaffilmark{1},
X.~Shu$\!$\altaffilmark{2},
S.~P.~Willner$\!$\altaffilmark{3},
M.~L.~N.~Ashby$\!$\altaffilmark{3},
J.-S.~Huang$\!$\altaffilmark{3,4,5},
A.~Fontana$\!$\altaffilmark{6},
A.~Dekel$\!$\altaffilmark{7},
E.~Daddi$\!$\altaffilmark{1},
H.~C.~Ferguson$\!$\altaffilmark{8},
J.~Dunlop$\!$\altaffilmark{9},
L.~Ciesla$\!$\altaffilmark{1},
A.~M.~Koekemoer$\!$\altaffilmark{8},
M.~Giavalisco$\!$\altaffilmark{10},
K.~Boutsia$\!$\altaffilmark{6},
S.~Finkelstein$\!$\altaffilmark{11},
S.~Juneau$\!$\altaffilmark{1},
G.~Barro$\!$\altaffilmark{12},
D.~C.~.Koo$\!$\altaffilmark{12},
M.~J.~Micha{\l}owski$\!$\altaffilmark{9},
G.~Orellana$\!$\altaffilmark{13},
Y.~Lu$\!$\altaffilmark{14},
M.~Castellano$\!$\altaffilmark{6},
N.~Bourne$\!$\altaffilmark{9},
F.~Buitrago$\!$\altaffilmark{9},
P.~Santini$\!$\altaffilmark{6},
S.~M.~Faber$\!$\altaffilmark{12},
N.~Hathi$\!$\altaffilmark{15},
R.~A.~Lucas$\!$\altaffilmark{8},
P.~G.~P\'erez-Gonz\'alez\altaffilmark{16}
}

\altaffiltext{1}{Laboratoire AIM-Paris-Saclay, CEA/DSM/Irfu, F-91191 Gif-sur-Yvette, France}
\altaffiltext{2}{Department of Physics, Anhui Normal University, Wuhu, Anhui, 241000, China}
\altaffiltext{3}{Harvard-Smithsonian Center for Astrophysics, 60 Garden St., Cambridge, MA 02138, USA}
\altaffiltext{4}{National Astronomical Observatories of China, Chinese Academy of Sciences, Beijing 100012, China}
\altaffiltext{5}{China-Chile Joint Center for Astronomy, Chinese Academy of Sciences, Camino El Observatorio, 1515, Las Condes, Santiago, Chile}
\altaffiltext{6}{INAF-Osservatorio Astronomico di Roma, Via Frascati 33, 00040 Monte Porzio Catone, Italy}
\altaffiltext{7}{Center for Astrophysics and Planetary Science, Racah Institute of Physics, The Hebrew University, Jerusalem, 91904, Israel}
\altaffiltext{8}{Space Telescope Science Institute, 3700 San Martin Drive, Baltimore, MD 21218, USA}
\altaffiltext{9}{SUPA, Institute for Astronomy, University of Edinburgh, Royal Observatory, Edinburgh, EH9 3HJ, U.K.}
\altaffiltext{10}{Department of Astronomy, University of Massachusetts, Amherst, MA, USA}
\altaffiltext{11}{Department of Astronomy, The University of Texas at Austin, Austin, TX 78712, USA}
\altaffiltext{12}{University of California Observatories/Lick Observatory, University of California, Santa Cruz, CA 95064, USA}
\altaffiltext{13}{Astronomy Department of Universidad de Concepci\'on, Concepci\'on, Chile}
\altaffiltext{14}{Observatories of the Carnegie Institution for Science, 813 Santa Barbara Street, Pasadena, CA 91101, USA}
\altaffiltext{15}{Aix Marseille Universite, CNRS, LAM (Laboratoire d'Astrophysique de Marseille) UMR 7326, 13388, Marseille, France}
\altaffiltext{16}{Departamento de Astrof\'{\i}sica, Facultad de CC.  F\'{\i}sicas, Universidad Complutense de Madrid, E-28040 Madrid, Spain}
\email{tao.wang@cea.fr}
\begin{abstract}

  We introduce a new color-selection technique to identify
    high-redshift, massive galaxies that are systematically missed by
    Lyman-break selection. The new selection is based on
    the $H_{160}$ ($H$) and Infrared Array Camera
    (IRAC) 4.5~\micron\ bands, specifically $H - [4.5] > 2.25$~mag.
    These galaxies, dubbed ``HIEROs'', include 
    two major populations that can be 
    separated with an additional $J - H$ color.  The populations are massive
    and dusty star-forming galaxies at $z > 3$  ($JH\text{-}blue$)
    and extremely dusty galaxies
    at $z \lesssim 3$ ($JH\text{-}red$).  The 350 arcmin$^{2}$
  of the GOODS-North and GOODS-South fields with the deepest
  {\it HST}/WFC3 near-infrared and IRAC data contain as many as 285 
  HIEROs down to $[4.5] < 24$~mag.  Inclusion of the most extreme
  HIEROs,  not even detected in the $H$ band, makes this
  selection particularly complete for the identification of
  massive high-redshift galaxies. We focus here primarily on $JH\text{-}blue$ ($z > 3$) HIEROs, which
  have a median photometric redshift $\langle z\rangle \sim 4.4$ and stellar mass
  $M_{*} \sim 10^{10.6}$~\Msol\ and are much fainter in the
  rest-frame UV than similarly massive Lyman-break galaxies (LBGs).  Their star formation
  rates (SFRs), derived from their stacked infrared spectral energy distributions, reach
  $\sim$240~\Msol~yr$^{-1}$ leading to a specific SFR, ${\rm sSFR} \equiv
  {\rm SFR}/M_{*} \sim4.2$~Gyr$^{-1}$, suggesting that the sSFRs
  for massive galaxies continue to grow at $z > 2$ but at a lower
  growth rate than from $z=0$ to $z=2$. With a median half-light
  radius of 2~kpc, including ${\sim}20\%$ as compact as quiescent
  galaxies at similar redshifts,  $JH\text{-}blue$ HIEROs represent perfect star-forming
  progenitors of the most massive ($M_{*} \gtrsim 10^{11.2}$~\Msol)
  compact quiescent galaxies at $z \sim 3$ and have the
  right number density. HIEROs make up ${\sim} 60\%$ of all galaxies
  with $M_{*} > 10^{10.5}$~\Msol\ identified at $z > 3$ from their
  photometric redshifts. This is five times more than LBGs with nearly
  no overlap between the two populations. While HIEROs make up
  15--25\% of the total SFR density at $z \sim 4$--5, they
  completely dominate the SFR density taking place in $M_{*} >
  10^{10.5}$~\Msol\ galaxies, and HIEROs are therefore crucial to
  understanding the very early phase of massive galaxy formation.
\end{abstract}

\section{Introduction}

Our current understanding of the cosmic star formation history at $z
\gtrsim 3$ is mostly based on studies of UV-selected samples, e.g.,
Lyman-break galaxies (LBGs). Yet the LBG selection is known to be
biased significantly against massive galaxies ($M_{*} \gtrsim 10^{11}$~\Msol,
because of both the relative faintness and redder UV slopes
for massive galaxies \citep{vandokkum:2006, Bian:2013}.
On the other hand, studies of the stellar mass function based on
photometric-redshift-selected galaxies from CANDELS reveals a
deficiency of galaxies at the massive end at $z \sim 4$
\citep{Grazian:2015}, suggesting that even the deepest near-infrared-selected
sample misses some massive
(dusty) galaxies. Understanding the selection bias of  different samples
and studying star formation in  more complete massive-galaxy
populations at $z \gtrsim 3$ are key to mapping the full cosmic star
formation history as well as understanding the very early phases of
massive galaxy formation.

In the last decade, a variety of massive, non-UV-selected galaxy
populations have been revealed and spectroscopically confirmed at $z
\gtrsim 3$, e.g., high-redshift submillimeter galaxies
\citep{Dunlop:2004,Chapman:2005,Younger:2007,Daddi:2009a,Riechers:2010,Capak:2011,Vieira:2013,HuangJ:2014},
quasars \citep{WangR:2013,Wagg:2014}, radio galaxies
\citep{Seymour:2007}, and red SPIRE sources
\citep{Casey:2012,Riechers:2013}. These galaxy populations usually
exhibit enormous star formation rates but low space densities
($\lesssim 10^{-5} - 10^{-6}$ Mpc$^{-3}$) and are unlikely to
represent the major population of massive galaxies. On the other
hand, recent observations reveal 
massive galaxies already formed and quenched at $z \sim 2.5$ with
space densities of ${\sim} 10^{-4}$~Mpc$^{-3}$
\citep{Daddi:2004,Fontana:2009,Brammer:2011,Muzzin:2013b,Ilbert:2013,vanderWel:2014}. 
This requires the existence of a significant population of
massive (star-forming) galaxies at $z \gtrsim 3$, as also predicted
in cosmological simulations~\citep[see., e.g.,
][]{Dekel:2014,Zolotov:2015,Feldmann:2015}. However, an efficient way
to identify the bulk population of $z \gtrsim 3$ massive galaxies is
still lacking.

One of the most efficient ways of selecting massive galaxies is color
selection, a method that relies on the strong spectral breaks
prevalent in massive galaxies. This technique allows a rather clean
selection of galaxies at certain redshifts and enables fair
comparisons between different studies. A number of
color-selection methods have been proposed to select $z > 3$
galaxies, making use of either the Balmer break \citep{Nayyeri:2014}
or the 1.6~$\mu$m stellar bump \citep{Mancini:2009} as redshift
indicators. One concern of these selection methods is contamination
from dusty galaxies and AGNs at lower redshifts ($z \sim 2-3$):
extremely dusty galaxies or AGNs at lower redshifts present similarly
red colors (due to attenuation or intrinsic red SEDs ) which mimic
those of the massive galaxies at high redshift (due to Balmer or the
1.6~$\mu$m break). This is similar to what we have learned from
color-selection techniques at $z\sim 2$, e.g., Distant Red Galaxies
\citep{Franx:2003}, IRAC-selected Extremely Red Objects
\citep{Yanh:2004,WangT:2012}, and $K_{s}$ and IRAC-selected Extremely
Red Objects~(Wang et al. 2012)\nocite{Wangw:2012}. How to separate
true high-redshift ($z \gtrsim 3$) from these low-redshift
contaminants remains a challenge for color-selection
techniques. In practice, to recover the bulk populations with
color-selection techniques, we need a large range of NIR to
mid-infrared (MIR) colors. This requires deep NIR and MIR imaging,
which has now become available thanks to the recently completed
Cosmic Assembly Near-infrared Deep Extragalactic Legacy Survey
(CANDELS; \citealt{Koekemoer:2011,Grogin:2011}), the {\it Spitzer} Extended
Deep Survey (SEDS; \citealt{Ashby:2013}) and the S-CANDELS
survey~\citep{Ashby:2015}.

\begin{figure*}[!htb] 
\centering
\includegraphics[trim=-20 0 0
-20,clip,angle=270,width=0.45\linewidth]{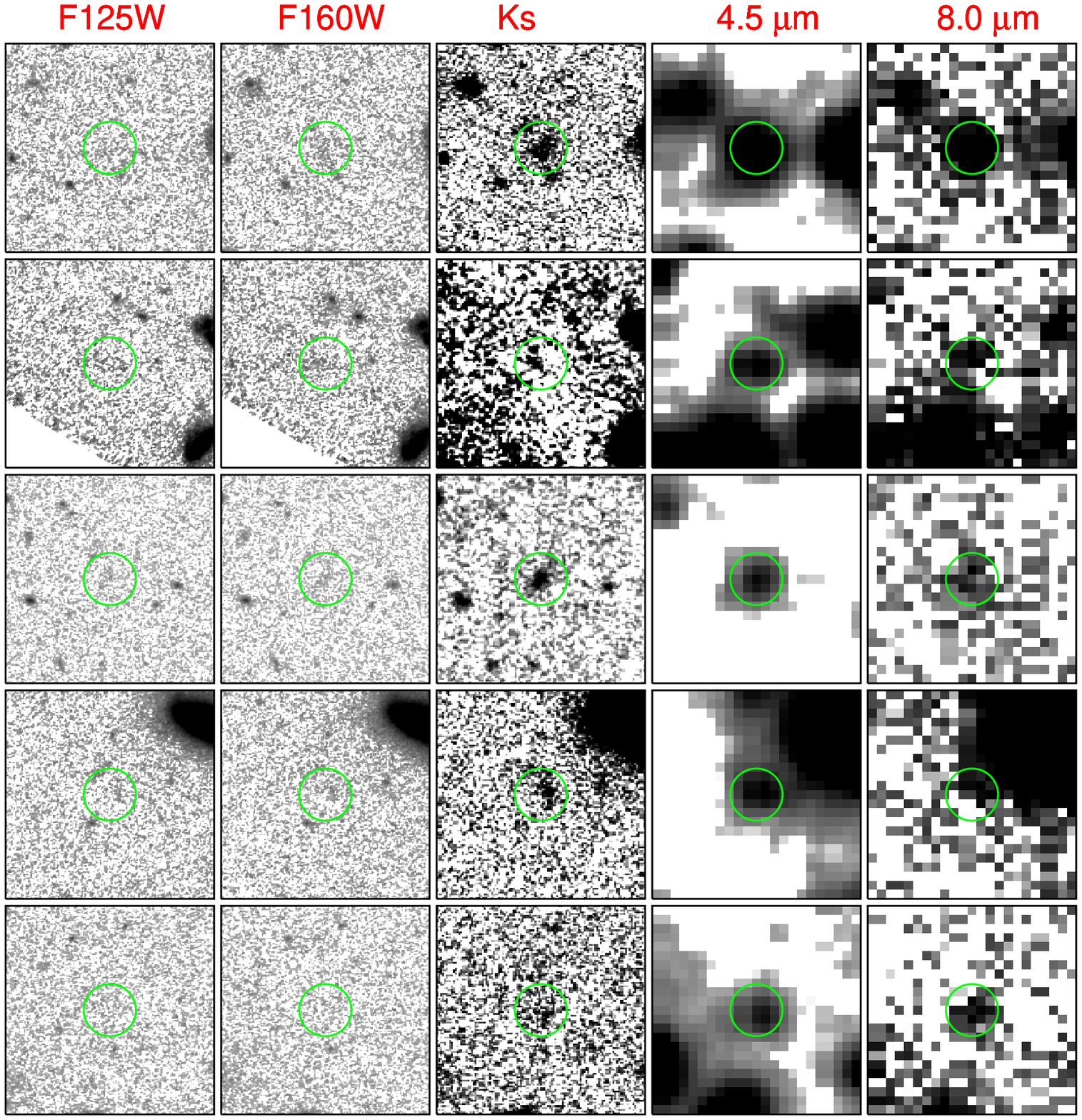}
\includegraphics[trim=-20 0 0
-20,clip,angle=270,width=0.45\linewidth]{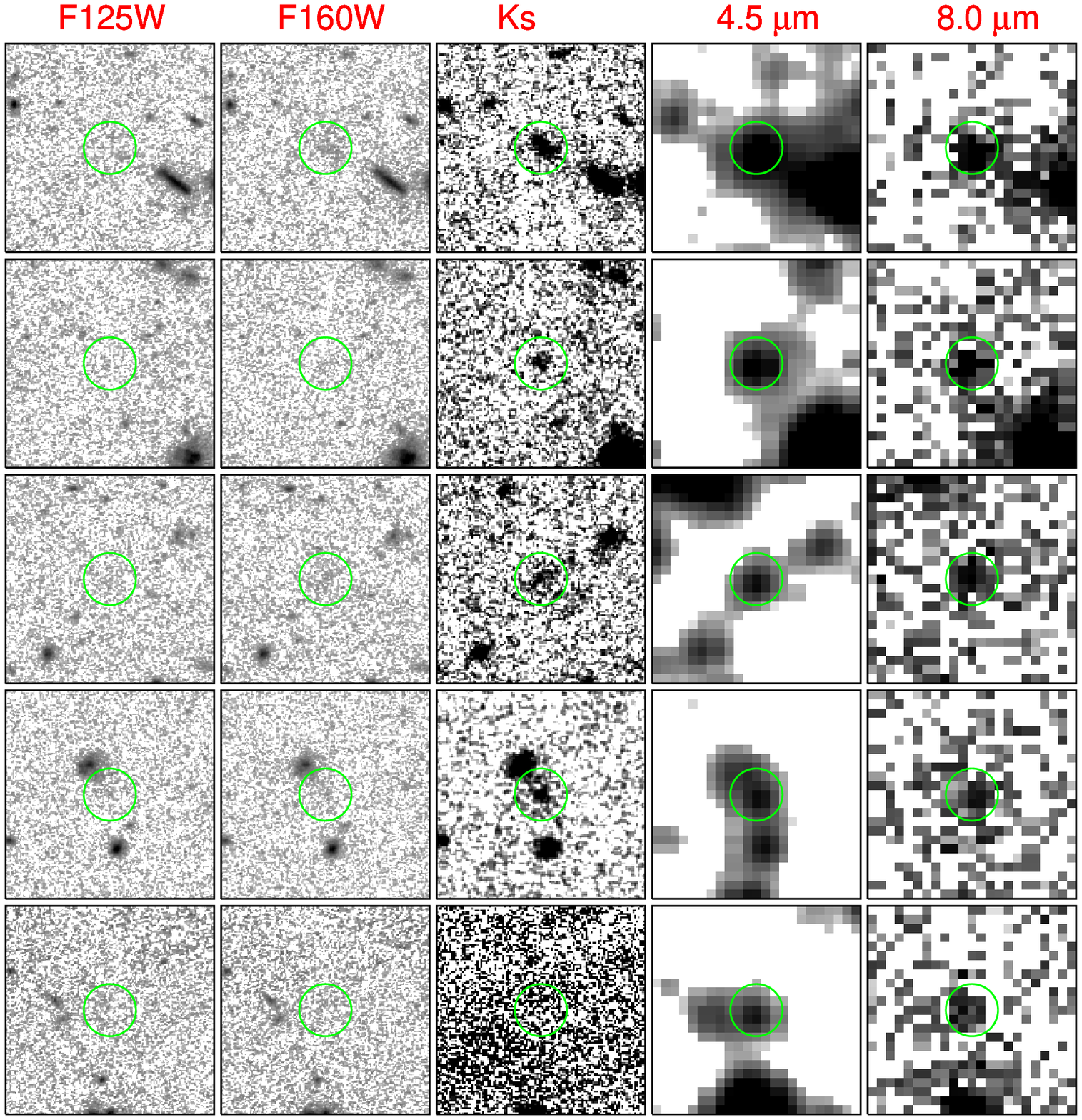}

\caption{Negative stamp images of example $H$-dropouts in GOODS-South in the
  F125W, F160W, $K_{s}$, 4.5~$\mu$m, and 8.0~$\mu$m
  bands. These $H$-dropouts are detected at IRAC 3.6 and
  4.5~$\mu$m with $[4.5] < 24$ yet with no $H$-band counterparts within
    2\arcsec\ after crossmatching with the CANDELS $H$-selected
    catalog. The size of each stamp image is $12\arcsec \times
  12\arcsec$.
\label{fig:stamp_Hdropout}}
\end{figure*}

A number of recent studies attempt to provide mass-selected samples
of galaxies at $z > 3$ based on photometric redshifts~\citep[see,
e.g., ][]{Fontana:2006,PerezGonzalez:2008,Spitler:2014,Straatman:2014,Marchesini:2014,Pannella:2015,Schreiber:2015}.
However, the quality of photometric redshifts of galaxies at $z
\gtrsim 3$ is quite uncertain due to the lack of large training
samples with spectroscopic redshifts. Although photometric redshifts
are believed to be more reliable in deep fields, e.g., in the HUDF
and GOODS fields, massive galaxies are rare, and larger fields are
needed to obtain statistically significant results. In this case,
color selection that requires only a few bands with high-quality
photometry has the advantage of efficiently identifying large samples
of certain populations of galaxies. Most existing studies
are based on near-infrared-selected samples, mostly $H$ band, which
probes the rest-frame UV. As a consequence, they may miss some of the
most massive and/or dustiest galaxies, as illustrated by
\cite{HuangJ:2011} \citep[see also, e.g., ][]{Caputi:2012}, who
reported a sample of galaxies that are bright in the IRAC bands yet
undetected in deep {\it HST }/WFC3 $H$-band, i.e., $H$-dropouts. Photometric
redshifts indicate that these galaxies are likely $z \gtrsim 4$
massive and dusty galaxies. Similarly, Wang et al. (2012) identified
a population of massive galaxies that are bright in IRAC but not
detected in the $K_{s}$-band. These are believed to be massive and
dusty galaxies at high redshift. Determining the number density and
star formation properties of these galaxies is essential to obtain a
complete view of both galaxy stellar mass functions and cosmic star
formation rate densities at high redshifts.

This paper presents a new color selection of Extremely Red
Objects with $H$ and IRAC colors (HIEROs, $H - [4.5] > 2.25$). This
color-selection technique is designed to select massive galaxies at
that are systematically missed by the Lyman-break selection
technique. We further show that combining $J - H$ colors
breaks the degeneracy between redshift and attenuation and enables a
clean selection of $z > 3$ galaxies.  We use the deep {\it HST }/WFC3
$H$-band
imaging from CANDELS and IRAC imaging from the SEDS survey and select
both $H$-detected and $H$-dropout HIEROs. Utilizing the exquisite
multi-wavelength dataset in the GOODS fields, we explore both their
star formation and structural properties.

This paper is organized as follows. We describe the data and the
selection of HIEROs in Section~\ref{Sec:data}. Photometric redshift
and stellar population analysis are presented in
Sections~\ref{Sec:redshifts} and ~\ref{Sec:classification},
respectively. Section~\ref{Sec:SF} explores star formation
properties and  Section~\ref{Sec:structure} structural properties
of HIEROs. Section~\ref{Sec:discussion} discusses the
completeness of the HIERO criteria in selecting $z > 3$ massive
galaxies, and Section~\ref{Sec:conclusion} summarizes.
Throughout the paper, we assume
cosmological parameters of $H_{0}$ = 70 km s$^{-1}$ Mpc$^{-1}$,
$\Omega_{M}$ = 0.3, and $\Omega_{\Lambda}$ = 0.7. All magnitudes are
in the AB system, where an AB magnitude is defined as ${\rm AB}
\equiv -2.5 \log({\rm flux\ density\ in\ \mu Jy}) +23.9$.

\section{Multi-Wavelength Data sets and sample selection}
\label{Sec:data}

\begin{table*}[!htb]\centering
\tablecolumns{10}
\ra{1.3}
\caption{Measured properties of all $H$-dropouts\label{tab:Hdropout_flux}}
\begin{tabular}{@{}cccccccccc@{}}
\toprule
  RA\footnote{Positions are from the IRAC 3.6 and 4.5~$\mu$m selected
  catalogs based on the SEDS survey \citep{Ashby:2013}} & Dec &
F160W & $K_s$ & 3.6~$\mu$m & 4.5~$\mu$m & 5.8~$\mu$m & 8.0~$\mu$m &
$z_{\rm phot}$ & Notes\footnote{have X-ray or 24~$\mu$m counterparts within 2\arcsec~after crossmatching with X-ray and 24~$\mu$m catalogs. These galaxies are most likely AGNs if they are indeed at $z > 3$. But also we note that their photometric redshifts estimate with FAST is less reliable.} \\
 \multicolumn{2}{c}{J2000}  & [$\mu$Jy] & [$\mu$Jy] & [$\mu$Jy] & [$\mu$Jy] &  [$\mu$Jy] & [$\mu$Jy] &  & \\\hline

GOODS-South\\

 53.19988 & -27.90455 & 0.32$\pm$0.03  & 0.85$\pm$0.06  &  3.56$\pm$0.42 & 5.50$\pm$0.53 & 8.55$\pm$0.56 & 11.85$\pm$0.65 & 4.78 & 24~$\mu$m\\
 53.12757 & -27.70675 & 0.31$\pm$0.02  &  0.93$\pm$0.06  &   2.83$\pm$0.36 & 3.84$\pm$0.41 & 4.87$\pm$0.52 & 6.07$\pm$0.58 & 4.62 & --\\
 53.04768 & -27.86865 & 0.26$\pm$0.04  & 0.39$\pm$0.14 & 1.56$\pm$0.23 & 1.96$\pm$0.27 & 3.54$\pm$1.02 & 3.22$\pm$1.07 & 5.17 & --\\
 53.08476 & -27.70800 & 0.05$\pm$0.02  & 0.15$\pm$0.06 & 1.24$\pm$0.20 & 2.25$\pm$0.29 & 3.27$\pm$0.56 & 4.24$\pm$0.58 & 5.26 & X-ray \\
 53.11912 & -27.81396 & 0.12$\pm$0.01  &  0.46$\pm$0.04 & 1.18$\pm$0.19 & 1.57$\pm$0.22 & 1.97$\pm$0.34 & 3.08$\pm$0.36 & 4.16 & -- \\
 53.06091 & -27.71833 & 0.26$\pm$0.02  & 0.48$\pm$0.06  & 1.04$\pm$0.16 & 1.47$\pm$0.20 & 1.02$\pm$0.69 & 2.40$\pm$0.67 & 3.19 & -- \\
 53.13463 & -27.90748 & 0.04$\pm$0.03 & 0.42$\pm$0.06 & 1.64$\pm$0.23 & 1.72$\pm$0.24 & 3.14$\pm$0.58 & 5.88$\pm$0.70 & 4.16 & --\\
 53.19654 & -27.75699 & 0.13$\pm$0.02  & 0.62$\pm$0.05 & 0.91$\pm$0.15 & 1.36$\pm$0.20 & 3.24$\pm$0.44 & 2.63$\pm$0.52 & 3.96 & --\\
 53.13270 & -27.72021 & 0.09$\pm$0.02  & 0.29$\pm$0.06 &1.03$\pm$0.16 & 1.33$\pm$0.20 & 0.68$\pm$0.84 & 2.40$\pm$0.65 & 4.68 & --\\
 53.02079 & -27.69903 & 0.05$\pm$0.03  &  0.17$\pm$0.50 &0.79$\pm$0.14 & 1.19$\pm$0.18 & 0.00$\pm$0.02   & 2.32$\pm$0.65 & 4.58 & --\\
  GOODS-North\\
  
    189.30783  &    62.30737 & 0.24$\pm$0.02 & 1.03$\pm$0.26 & 3.60$\pm$0.46 & 6.03$\pm$0.58 & 8.74$\pm$0.47 &15.59$\pm$0.54&3.99 & 24~$\mu$m,X-ray \\
    189.18352  &    62.32741 & 0.25$\pm$0.01 & -- & 2.68$\pm$0.37 & 4.37$\pm$0.47 & 4.76$\pm$0.75 & 9.67$\pm$0.66& 7.00 & 24~$\mu$m\\
    189.42834  &    62.26596 & 0.10$\pm$0.03 & 0.46$\pm$0.19 & 1.27$\pm$0.22 & 1.22$\pm$0.18 & 2.03$\pm$0.62 & 5.14$\pm$0.55& 7.00 & 24~$\mu$m\\
    189.25689  &    62.25028 & 0.22$\pm$0.01 & 0.23$\pm$0.16 & 1.13$\pm$0.19 & 1.14$\pm$0.18 & 1.24$\pm$0.47 & 2.50$\pm$0.46& 6.95 & --\\
    189.39476  &    62.31691 & 0.17$\pm$0.02 & 0.26$\pm$0.17 & 1.05$\pm$0.18 & 1.20$\pm$0.19 & 1.86$\pm$0.48 & 2.85$\pm$0.60& 6.92 & --\\
    189.02396  &    62.22296 &0.02$\pm$0.05 & 0.20$\pm$0.16 & 0.95$\pm$0.17 & 1.28$\pm$0.20 & 1.59$\pm$0.52 & 2.57$\pm$0.63& 4.54 & --\\ 
\bottomrule
\end{tabular}

\end{table*}

\subsection{A combined F160W and IRAC 4.5$\mu$m-selected catalog}
\label{subsec:cat}
The GOODS-South and GOODS-North fields have been the target of some
of the deepest surveys ever conducted over a broad wavelength range
by space observatories and the foremost ground-based telescopes.  In
particular, the new HST/WFC3 near-infrared survey from CANDELS and
the {\it Spitzer}/IRAC mid-infrared survey from SEDS have significantly
improved measurements of galaxy properties at $z > 2$.  Here we
utilize the UV to mid-infrared multi-wavelength catalogs based on
detections in the $HST$/WFC3 F160W band from CANDELS for both
GOODS-South and GOODS-North as described by \cite{GuoY:2013} and
Barro et al. (in preparation), respectively.  Both fields include
deep photometry in the F435W, F606W, F775W, F814W, F850LP, F105W,
F125W, F140W, and F160W bands from $HST$ and 3.6, 4.5, 5.8, and
8.0~$\mu$m bands from {\it Spitzer}/IRAC \citep{Ashby:2013}. The
GOODS-South catalog also includes photometry in the $U$-band from
both CTIO/MOSAIC and VLT/VIMOS and $Ks$ -band imaging from the
Infrared Spectrometer and Array Camera (ISAAC) and the High Acuity
Wide field K-band Imager (HAWK-I) on the VLT \citep[HUGS survey,
][]{Fontana:2014}. Similarly, the GOODS-North catalog includes
photometry in the $U$-band from both KPNO and LBT and $Ks$-band imaging
from both the Multi-Object Infrared Camera and Spectrograph (MOIRCS)
and CFHT. Both fields reach 5$\sigma$ depth of $H\sim27.2$~mag and
are 80\% complete down to $H\sim26.2$.

\begin{figure*}[!htb] 
\centering
\includegraphics[trim= -20 -30 0  0,clip, width=0.49\linewidth]{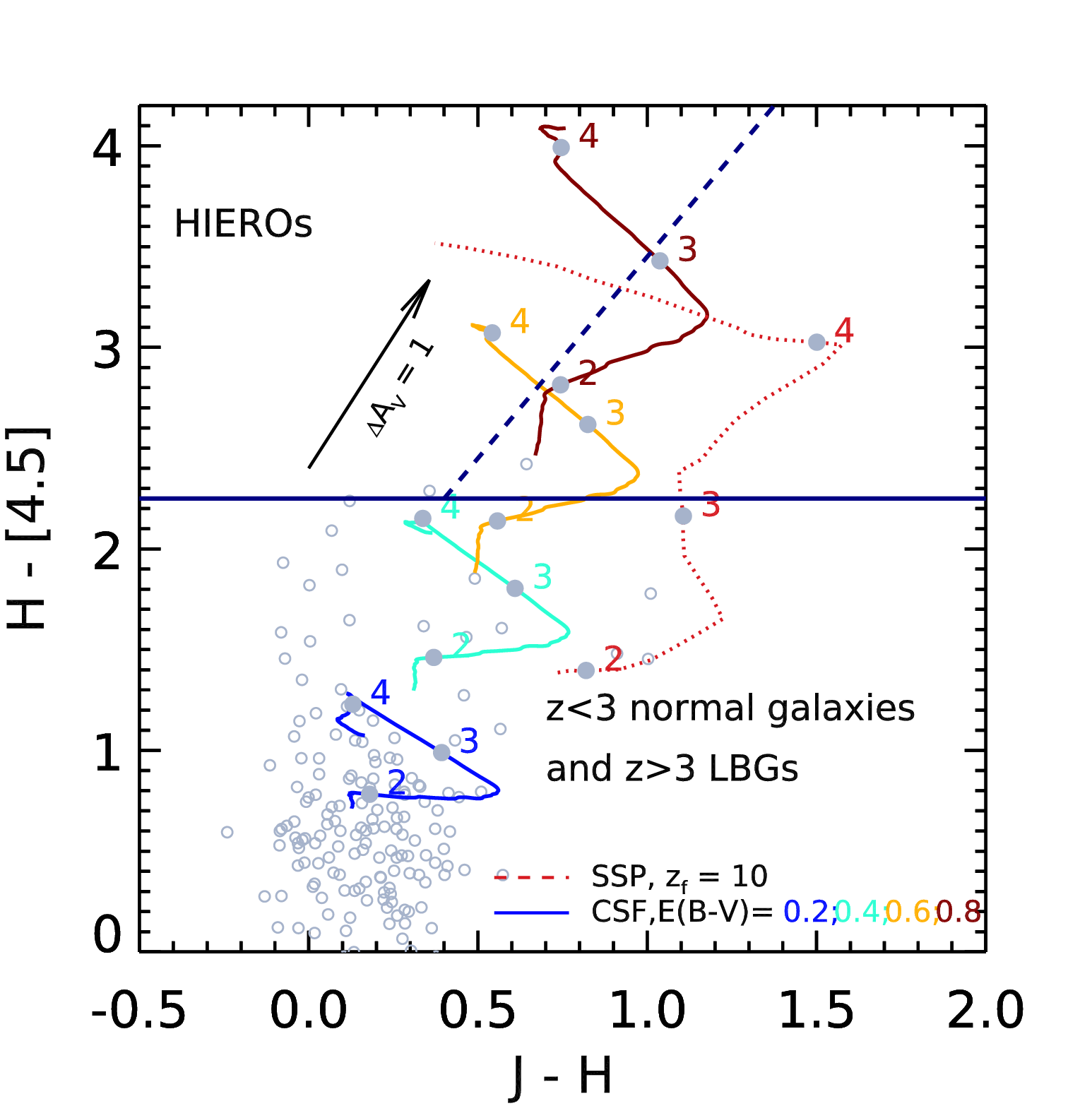}
\includegraphics[trim= -25 -20 0 20,clip,width=0.5\linewidth]{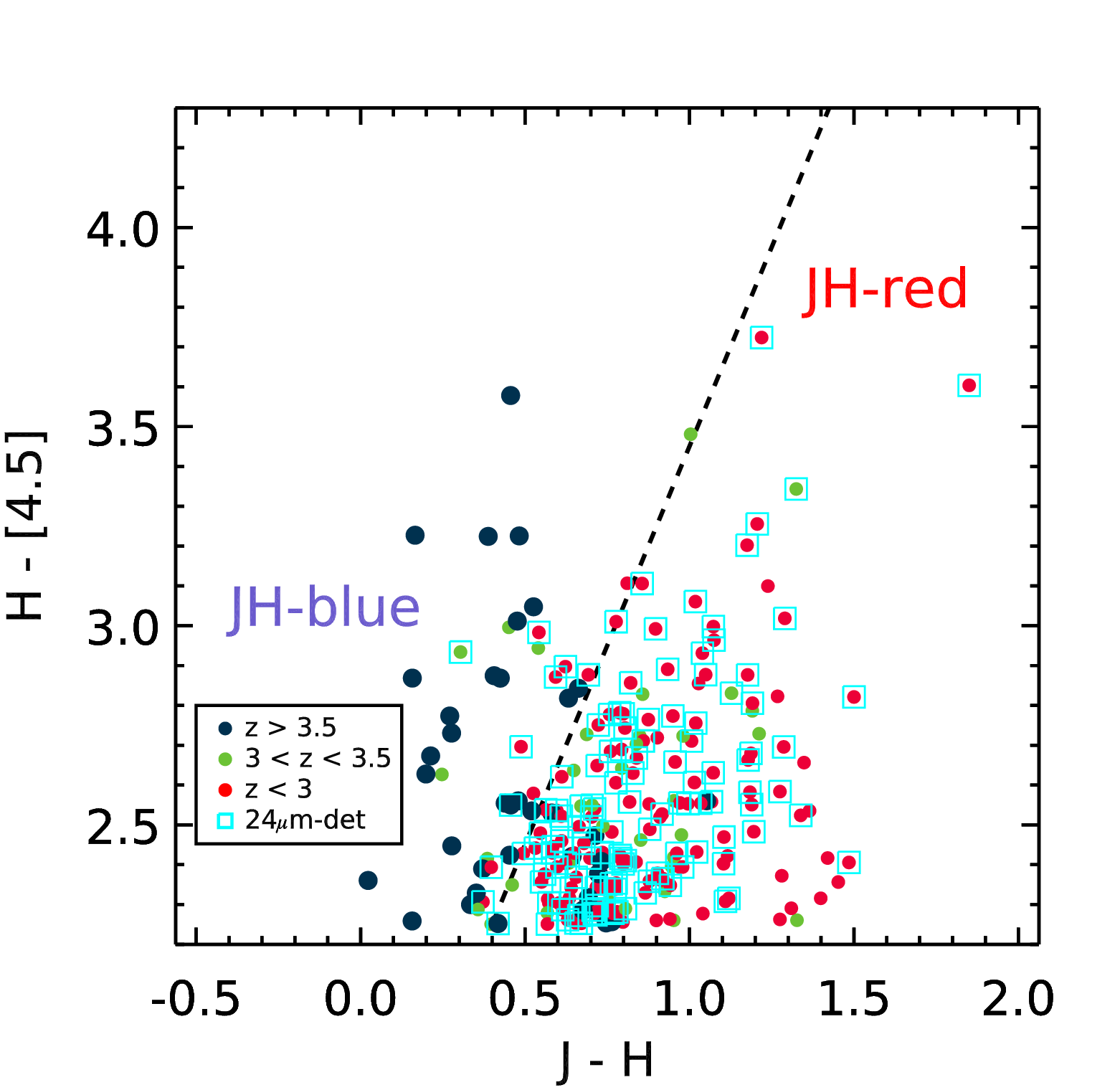}
\caption{\textbf{Left panel}: Color-color diagram for the HIERO
  selection based on the $H - [4.5]$ colors. Evolutionary tracks of a
  set of theoretical galaxy SED templates between $z = 5$ and $z = 2$
  are shown, including an instantaneous burst (SSP) model formed at
  $z = 10$ and a constant star formation model (CSF) of age 300~Myr
   with different levels of reddening. The solid horizontal line
  shows the HIERO selection criterion adopted.The diagonal dashed
  line separates pure $z > 3$ dusty galaxies from passive galaxies at
  $z > 3$ and extremely dusty galaxies at lower redshifts. Open
  circles denote galaxies with spectroscopic redshifts $z > 3$ in the
  two GOODS fields. These are mostly UV-bright galaxies with lower
  levels of attenuation, i.e., LBGs. \textbf{Right panel}: The
  distribution of HIEROs with detections ($>$5$\sigma$ ) in both $J$ and $H$ selected in the GOODS fields in the $H - [4.5]$ versus $J -
  H$ color-color diagram, color-coded by their
  redshifts. The diagonal dashed line separates $JH\text{-}blue$ and
  $JH\text{-}red$ HIEROs as given by Equations 1 and 2 (the same
  dashed line as shown in the left panel). Galaxies detected
    at 24~$\mu$m ($F_{24~\micron} > 30$~\uJy) are shown by cyan
    squares. Note that 24~$\mu$m-detected sources are prevalently star-forming galaxies at $z < 3$ and classified as
     $JH\text{-}red$ HIEROs, as expected. Galaxies not detected
  in the F125W ($J$) band are shown with their 3$\sigma$ upper limits.
  \label{fig:color_evolve}}
\end{figure*}

We also performed a systematic search for objects that are bright in
the IRAC bands but are missed in the $H$-selected catalog, i.e.,
$H$-dropouts. We crossmatched the CANDELS $H$-selected catalog with an
IRAC 3.6 and 4.5~$\mu$m selected catalog \citep{Ashby:2013} from the SEDS survey. The
SEDS survey covers the two GOODS fields to a depth of 26 AB mag
(3$\sigma$) at both 3.6 and 4.5~$\mu$m and is 80\% complete down to
[4.5] $\sim$24~mag. We first matched sources with $[4.5] < 24$~mag in
the SEDS catalog to the $H$-selected catalog and
identified those without $H$-band counterparts within a 2\arcsec\
radius. This 4.5~$\mu$m magnitude cut was applied to enable
sufficient color range to identify extremely red objects and also give a
complete 4.5~$\mu$m-selected sample. We then visually inspected the
IRAC images and excluded sources whose flux is likely contaminated by
bright neighbors as well as those falling on the edge of the F160W
image. We dubbed this catalog of IRAC sources with no $H$-band
counterparts the ``H-dropout''
catalog. Figure~\ref{fig:stamp_Hdropout} shows examples of
$H$-dropouts identified in GOODS-South. With knowledge of their
positions, some of these $H$-dropouts
are marginally detected in the $H$~band but exhibit extended profiles
and are unidentifiable as real sources without that prior
knowledge. We measured aperture magnitudes in $H$ and $K_{s}$-bands with a 1\arcsec~radius aperture
 at the position of their IRAC 3.6 and 4.5~$\mu$m counterparts, and then applied an aperture correction to get the total flux. 
 For IRAC 5.8 and 8.0~$\mu$m bands, we used a 1.2~\arcsec radius aperture plus aperture correction, 
 the same as that used for IRAC 3.6 and 4.5~$\mu$m bands (directly taken from \cite{Ashby:2013}).    
 Their measured flux densities across $HST$/WFC3 to 
IRAC bands are listed in Table~\ref{tab:Hdropout_flux}.

The combined F160W and IRAC 4.5~$\mu$m-selected catalog is not only
complete to [4.5] = 24~mag but also ensures that all the $H$-dropouts
have $H - [4.5] > 2.25$. As shown by \cite{GuoY:2013}, the agreement
of the IRAC 4.5~$\mu$m photometry between the CANDELS F160W-selected
catalog and the SEDS 3.6~$\mu$m and 4.5~$\mu$m selected catalog is
excellent for objects with $[4.5] < 24.5$~mag. The magnitude cut of
our selection, [4.5] = 24, is much brighter than the detection limit
in the SEDS survey, and therefore Eddington bias is also negligible
\citep{GuoY:2013}.

We searched for infrared and X-ray counterparts within a 2\arcsec\
radius for all the sources in both the $H$-selected and $H$-dropout
catalogs based on their $H$-band or IRAC positions. For infrared
counterparts, we employed the MIPS 24~$\mu$m-selected catalog of
\cite{Magnelli:2013}, which also includes 100~$\mu$m and 160~$\mu$m
photometry from the combination of PACS Evolutionary Probe (PEP;
\citealt{Lutz:2011}) and GOODS-$Herschel$ \citep{Elbaz:2011} key
programs. For X-ray counterparts, we used the 4~Ms catalog
\citep{Xue:2011} for GOODS-South and the 2~Ms catalog
\citep{Alexander:2003} for GOODS-North.

\subsection{Selection of  $z > 3$ massive galaxies}
\label{subsec:sample}

At $z > 3$, the Balmer/4000~\AA\ break shifts redward of the $H$ band
while the 4.5~$\mu$m band probes the rest-frame $J$ band. Thus both
quiescent galaxies with strong Balmer/4000~\AA\ breaks and dusty
galaxies with significant UV attenuation appear red in
$H-[4.5]$. Figure~\ref{fig:color_evolve} plots the evolution of $H -
[4.5]$ as a function of redshift for different sets of
templates. These templates are based on BC03 models, including a
non-evolving constant star formation (CSF) model computed for an age
of 300 Myr and various levels of reddening (using the
\cite{Calzetti:2000} extinction law and solar metallicity).   This
figure illustrates that an $H - [4.5] > 
2.25$ color cut can effectively select old or dusty galaxies at $z
\gtrsim 3$. On the other hand, most commonly used LBG selection
techniques \citep[e.g.,][]{Bouwens:2012} are designed specifically to
select young and less attenuated galaxies, with UV slope $\beta
\lesssim 0$ or equivalently $E(B-V) \lesssim 0.4-0.5$ for a young
star-forming galaxy. Therefore, the proposed red galaxy selection is
complementary to the LBG selection and is crucial for a complete
census of galaxy populations at $z \gtrsim 3$. As a further
illustration of this point from Figure~\ref{fig:color_evolve}, almost
none of the spectroscopically confirmed $z > 3$ galaxies (mostly
LBGs) have $H - [4.5] > 2.25$.  In the following sections, we refer
to galaxies with $H - [4.5] > 2.25$ as HIEROs.

Figure~\ref{fig:color_evolve} also reveals that while passive or
dusty galaxies at $z > 3$ are expected to be identified as HIEROs,
extremely dust-obscured galaxies ($E(B-V) \gtrsim 0.6$) at $2 < z <
3$ could also enter the HIERO selection. This is similar to other red
galaxy selection technique at lower redshifts. For instance,
\cite{Wuyts:2009b} found that 15\% of their distant red galaxy (DRGs,
\citealt{Franx:2003}) sample, which is intended to select galaxies at
$z > 2$, have spectroscopic redshifts $z < 2$ \citep[also see, e.g., ][]{Grazian:2007}. These low-redshift
DRGs are on average more obscured with $A_{V}$ higher by 1.2~mag than
the high-redshift DRGs. The situation is expected to be more serious
in selecting galaxies at $z > 3$ due to the prevalence of dusty
galaxies at $2 < z < 3$ \citep[see, e.g.,
][]{YanL:2007,Dey:2008,HuangJ:2009}.

\begin{figure}[!tb] 
\centering
\includegraphics[trim= 0 -20 0 0,clip,width=0.95\linewidth]{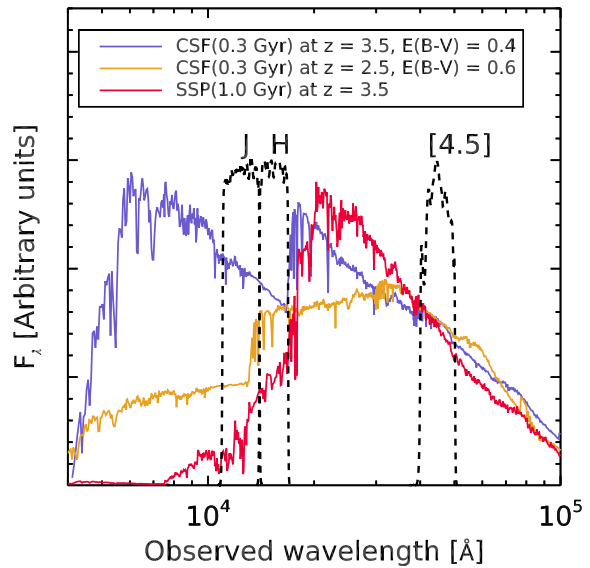}
\caption{Total transmission curves of the {\it HST}/WFC3 F125W, F160W, and
  IRAC 4.5~$\mu$m filters used to define the criteria for selecting
  $z > 3$ galaxies. Solid lines show theoretical galaxy templates for
  a $z = 3.5$ star-forming galaxy, a $z = 2.5$ extremely dusty
  galaxy, and a passive/old galaxy with an age of 1 Gyr at $z =
  3.5$. While the three galaxy templates all present similar red $H -
  [4.5]$ colors, an additional $J - H$ color can distinguish between
  them due to the differences in the position and strength of the
  4000~\AA~break as well as the UV slope.\label{fig:sed_comp}}
\end{figure}

\begin{figure*}[!htb] 
\centering
\includegraphics[trim= -40 -40 10 30,clip,width=0.49\linewidth]{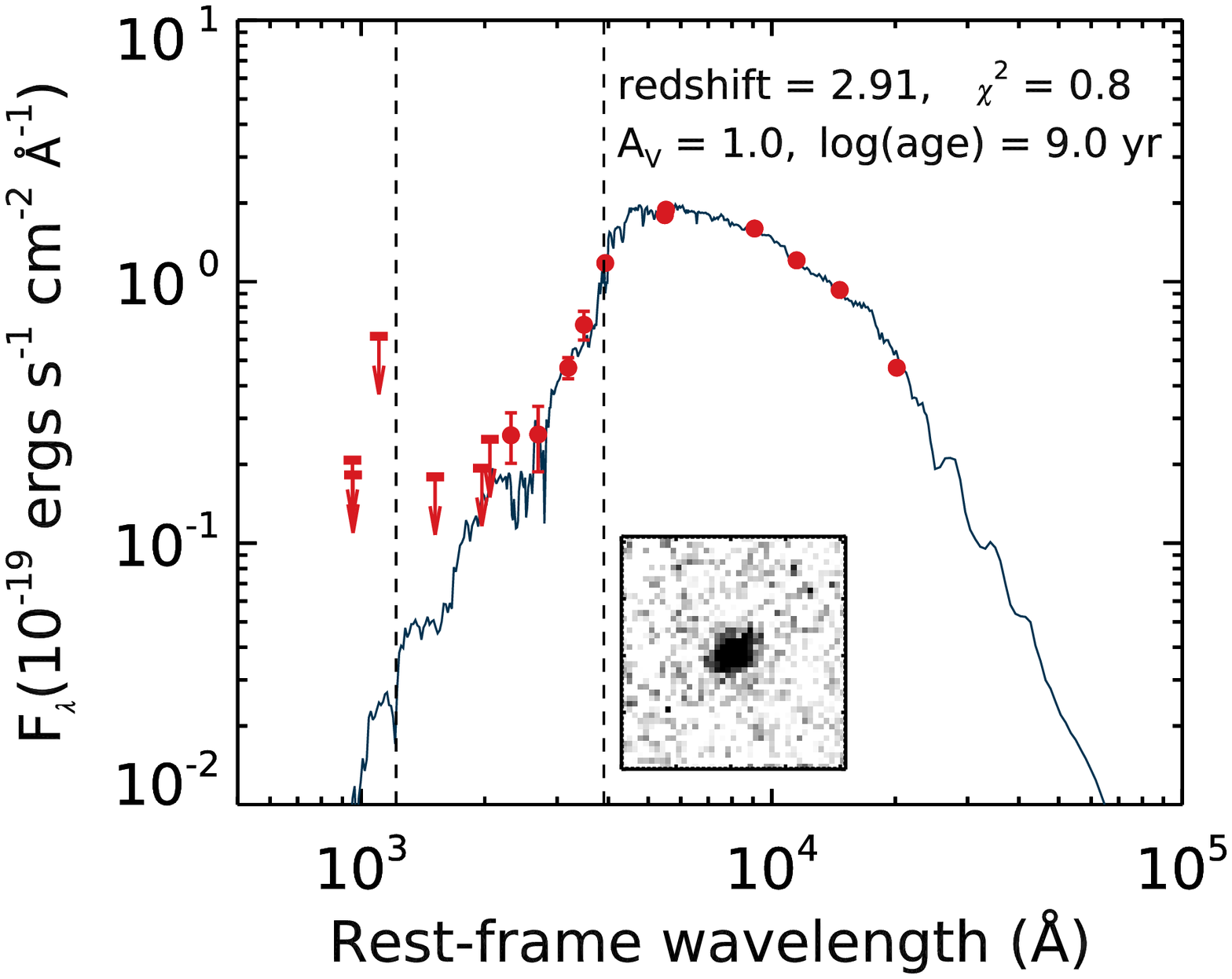}
\includegraphics[trim= -40 -40 10 30,clip,width=0.49\linewidth]{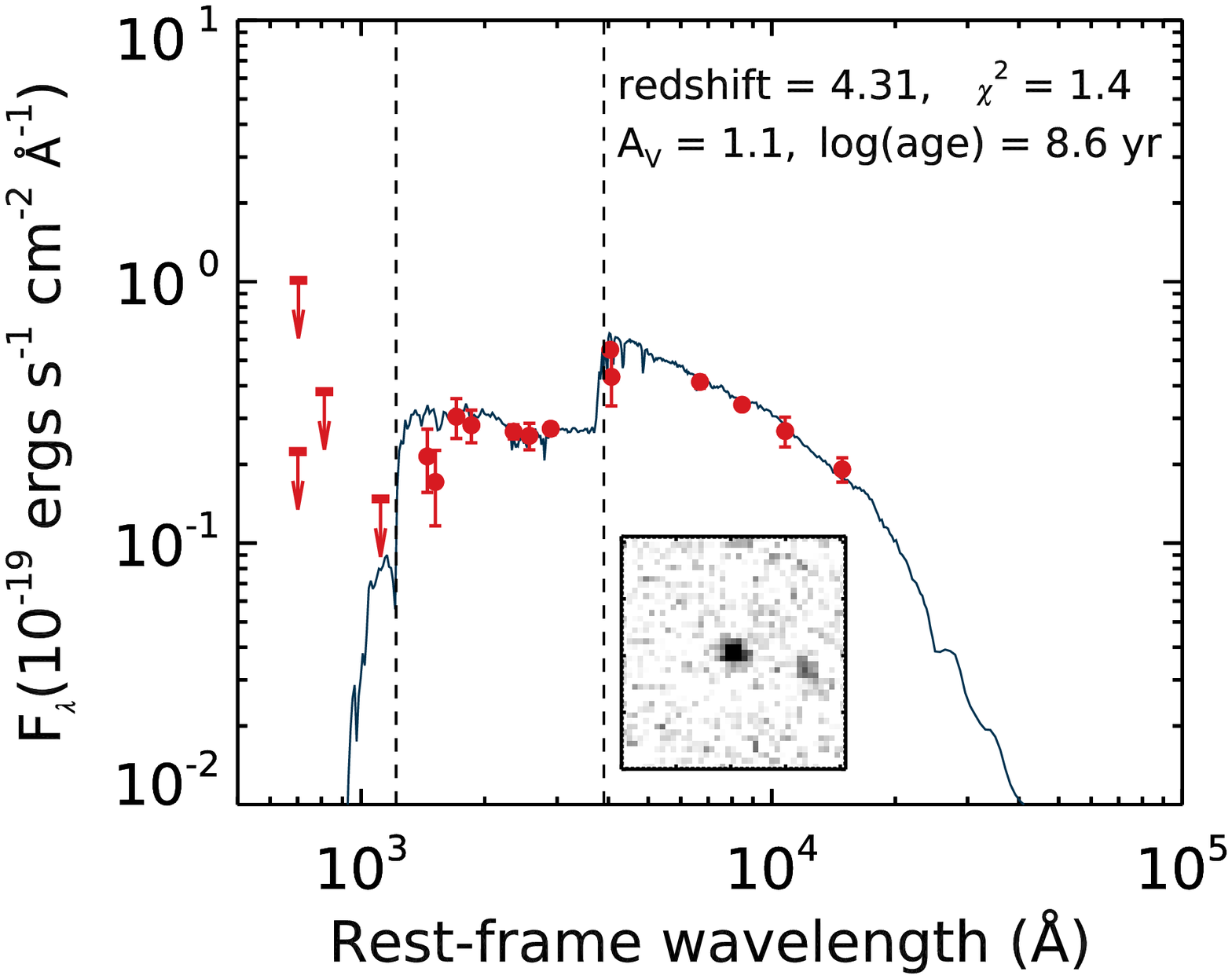}

\caption{Examples of SED-fitting results for a $JH\text{-}red$
  (\textbf{left}) and $JH\text{-}blue$ (\textbf{right}) HIERO using
  the full band photometry with FAST\null. Indicated upper limits are
  $3\sigma$.  Inset show
  F160W negative images with size  2.5$\arcsec
  \times 2.5 \arcsec$. \label{fig:fast_sed_fit}}
\end{figure*}

\begin{table*}\centering
\ra{1.3}
\caption{Number counts of HIEROs\label{tab:number_counts}}
\begin{tabular}{@{}lcccccc@{}}
\toprule
Sample        & Number  & Number  & Number & Number  & Number & Number\\
&  ($H$-detected) & ($H$-detected, clean\footnote{see Section~\ref{subset:clean}.}) & ($H$-dropouts) & (final sample) & ($F_{24~\micron} > 30$~\uJy)  & (X-ray\footnote{detected at 0.5--8~keV}) \\\hline

$JH\text{-}red$   &   243  & 206  & --    &  206  & 123 & 39\\
$JH\text{-}blue$   &   116  & 61   & 18    &  79    & 18  & 14 \\
All      &   359  & 267  & 18    &  285    &  141 & 53\\
\bottomrule
\end{tabular}
\end{table*}

\begin{figure}[!htb] 
\centering
\includegraphics[trim= -30 -30 0 0,clip,width=0.98\linewidth]{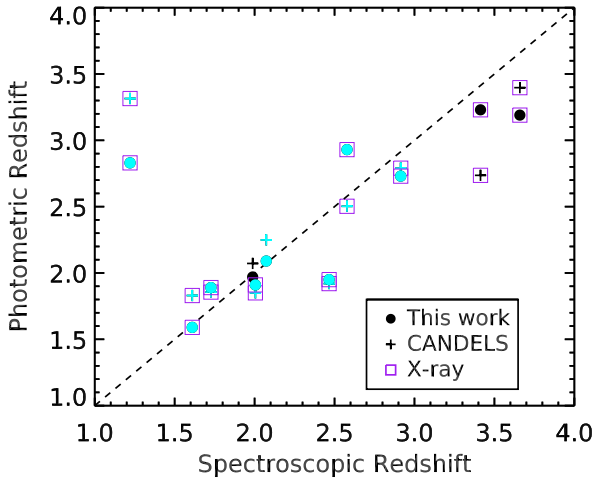}
\caption{Spectroscopic redshift versus photometric redshift for
  galaxies in our sample. 24~$\mu$m-detected galaxies are denoted in
  cyan while X-ray sources are denoted by purple open squares. Both
  photometric redshifts derived using $FAST$ in this work and those
  from CANDELS are presented. \label{fig:zspec_zphot}}
\end{figure}

\begin{figure*}[!htb] 
\centering
\includegraphics[trim= 0 0 10 20,clip,width=0.49\linewidth]{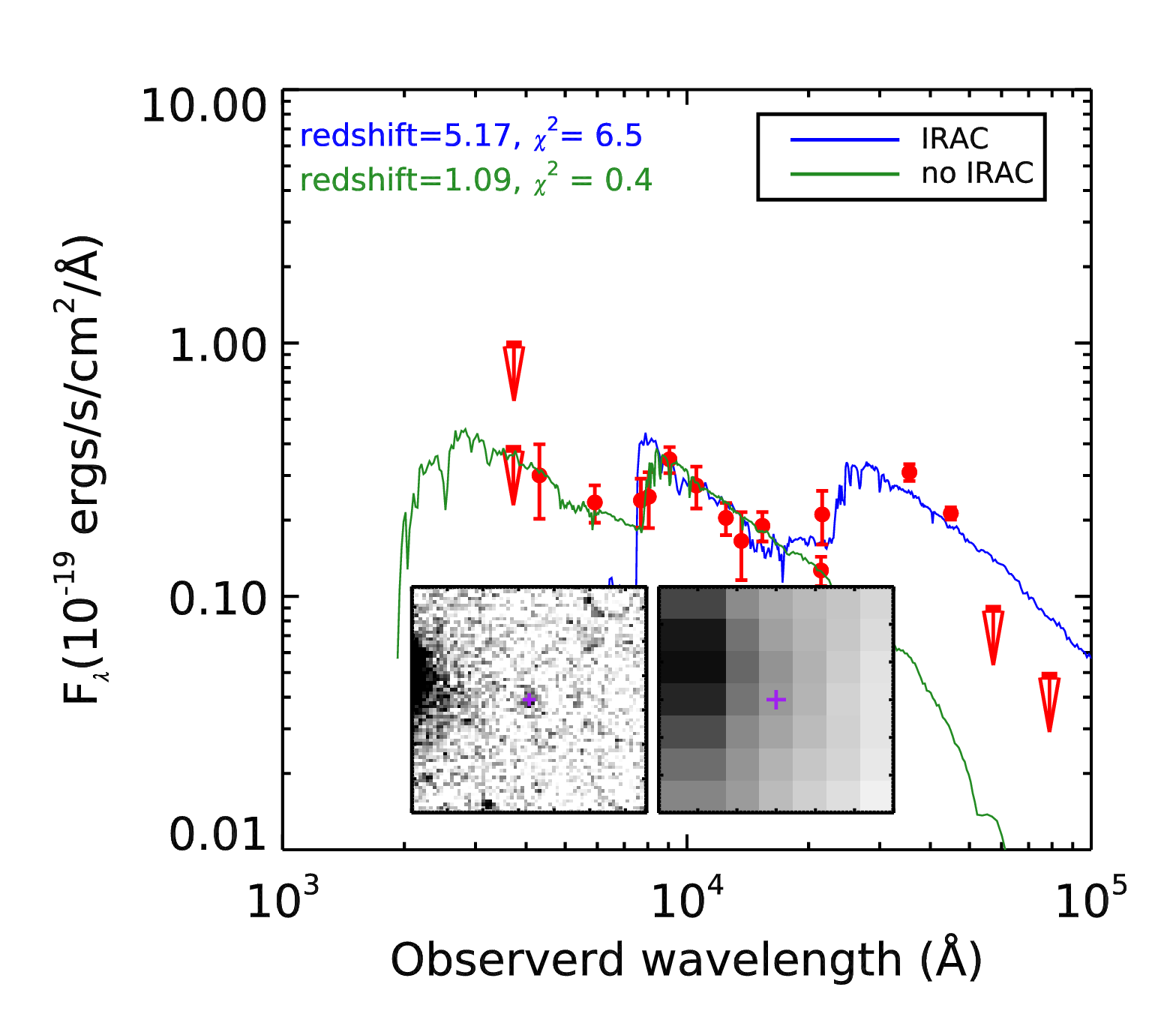}
\includegraphics[trim= 0 0 10 20,clip,width=0.49\linewidth]{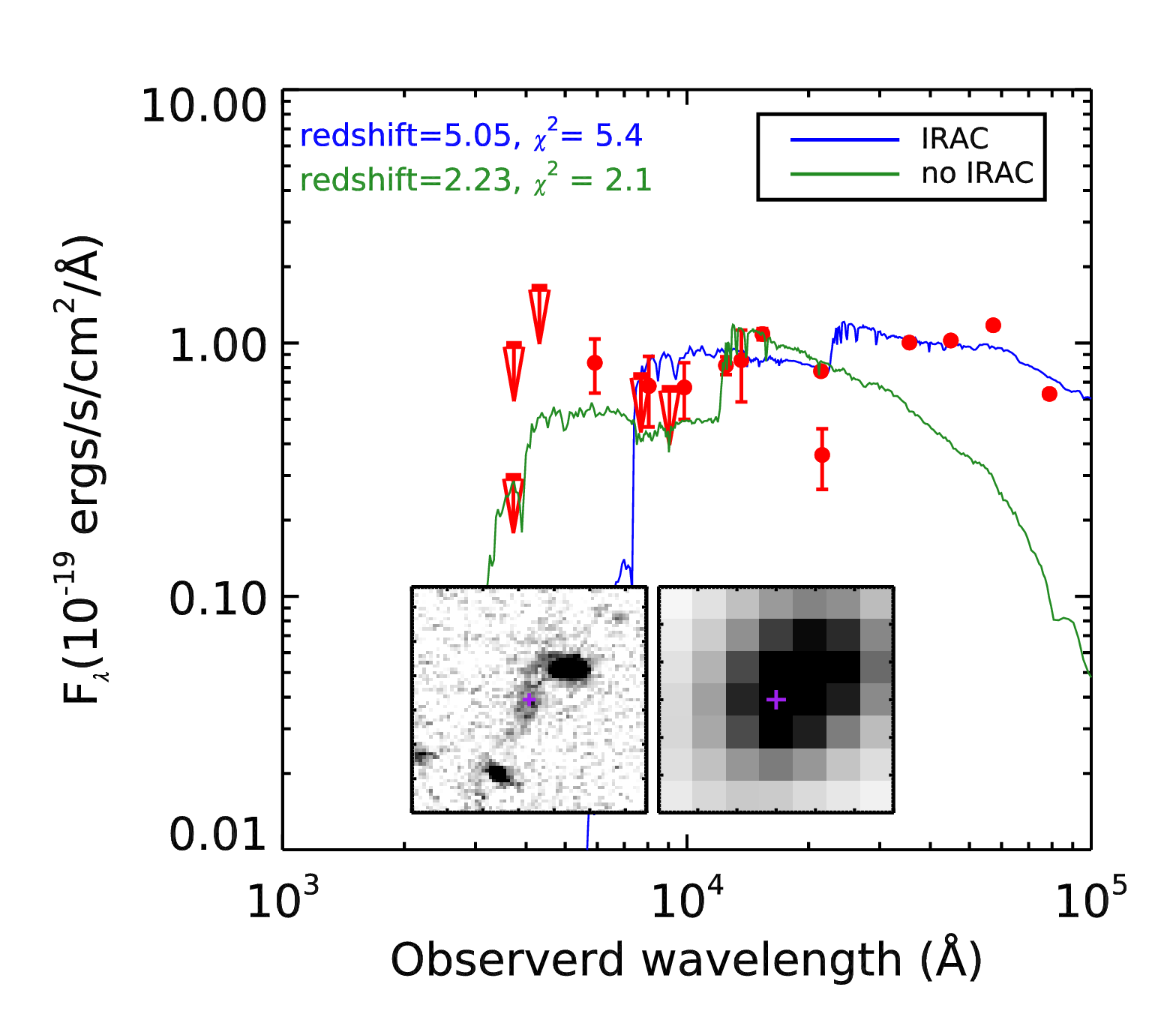}
\caption{Examples of SED fitting when including/excluding IRAC
  photometry. For each galaxy, from left to right the inner panels
  show the F160W and IRAC 4.5~$\mu$m images with the image size of
  4$\arcsec \times 4 \arcsec$. When including the IRAC photometry, a
  bad fit was achieved with $\chi^{2} \gtrsim 4$.  This suggests that
  most likely the IRAC photometry for these galaxies is contaminated
  by neighboring sources, as can also been seen from the stamp
  images. 
  \label{fig:fast_zfit_bad}}
\end{figure*}

To enable a cleaner selection of $z > 3$ galaxies, we  use
an additional $J - H$ color to separate $z > 3$ galaxies from
low-redshift contaminants, following the color track of theoretical
templates.  As shown in Figure~\ref{fig:sed_comp}, although a heavily
attenuated galaxy ($E(B-V) \gtrsim 0.4$) at $z < 3$ could have a
similar $H - [4.5]$ color as normal massive star-forming galaxies at
$z > 3$, the Balmer/4000~\AA~break falling between the $J$ and $H$
bands at $z \sim 2$--3 leads to a much redder $J - H$ (and also
$J - K_{s}$) color than for galaxies at $z > 3$. Similarly, passive galaxies at
$z > 3$ also have redder $J - H$ colors due to much redder rest UV slopes
than star-forming galaxies. Therefore,  an
additional $J - H$ color criterion  separates these different
populations and approaches a pure selection of $z > 3$ massive
(star-forming) galaxies.  Based on the color tracks of theoretical
models and the photometric redshifts of HIEROs (details of
photometric redshift determinations are discussed in
Section~\ref{Sec:redshifts}), we separate $z \gtrsim 3.5$
star-forming galaxies ($JH\text{-}blue$ HIEROs) from $z \sim 2-3$
contaminants ($JH\text{-}red$ HIEROs) as:
\begin{align}
JH\text{-}blue~\text{(high-}z\text{):} & ~ H - [4.5]  > 2 \times (J - H) + 1.45\\
JH\text{-}red~\text{(low-}z\text{):}  & ~ H - [4.5]  \leq 2 \times (J - H) + 1.45.
\end{align}
Because both $H - [4.5]$ and $J-H$ are poorly constrained for
$H$-dropouts, we classify all the $H$-dropouts as $JH\text{-}blue$ HIEROs
(the redder $H - [4.5]$ color of $H$-dropouts suggests that they are in
general at higher redshift).  In total, we identify 359 HIEROs (116
$JH\text{-}blue$ and 243 $JH\text{-}red$) in the two GOODS fields, 18
of which are $H$-dropouts. After examining the reliability of the color
measurements of independent sources with an approach described in
Section~\ref{subset:clean}, our final sample includes 285 (80\% of
the original sample) HIEROs.  We list respective fractions of the two
categories of HIEROs in Table~\ref{tab:number_counts}.

\cite{Caputi:2012} studied a sample of extremely red galaxies with $H
- [4.5] > 4$ in UDS. Only 15 galaxies in our sample present such
extremely red colors. While nearly all of these 15 galaxies have
$z_{\rm phot} > 3$, the majority of the massive $z > 3$ galaxies will be
missed by this extreme criterion. (A more extreme color cut leads to
fewer contaminants at lower redshifts but also to a lower completeness
in selecting high-redshift galaxies.) Similarly, \cite{Wangw:2012}
studied a sample of $K$- and IRAC-selected extremely red objects
(KIEROs, $K_{s} - [4.5] > 1.6$) in GOODS-North, aiming to identify
specifically dusty galaxies at $z > 2$. They showed that 
the majority of KIEROs are at $z \sim 2$--3.5. 46\% of our
HIERO sample satisfy the color criterion of KIEROs. Compared to both
previous studies, the advantage of our color selection is a
more complete sample of massive (including both passive and
star-forming) galaxies at $z > 3$ because our selection was specifically designed to
complement the LBG selection. This allows us to perform a complete
census of massive galaxy evolution at $z \gtrsim 3$. Moreover, with
the proposed $H - [4.5]$ and $J-H$ diagram, we are able to
distinguish high-redshift galaxies from low-redshift contaminants,
enabling a much cleaner (and also complete) selection of massive
galaxies compared to previous studies.

\section{Redshifts of HIEROs}
\label{Sec:redshifts}
\subsection{Photometric Redshifts}
The HIEROs are extremely faint at observed UV and visible wavelengths. Even though
the GOODS fields have been extensively covered by spectroscopic
observations, only 11 HIEROs have spectroscopic redshifts
\citep[][and references
therein]{Kajisawa:2010,Dahlen:2013,Hsu:2014}. Therefore we use
photometric redshifts to gain insight into their nature. The unique
features of these galaxies, e.g., many of them are consistent with
being extremely dusty with $A_{V}$ exceeding $3-4$, lead to concerns
that they may be not represented in the templates used by most
photometric redshift methods. Hence we used galaxy templates spanning
a larger parameter space particularly a larger range of $A_{V}$ to
determine their photometric redshifts. To this aim, we employed FAST
\citep{Kriek:2009a} to fit the full $U$-band to 8.0~$\mu$m photometry
for all galaxies in the HIERO sample. FAST also provides
self-consistent estimates of stellar masses. We constructed stellar
templates from the \cite{Bruzual:2003} (BC03, hereafter) stellar
population synthesis model with a \cite{Chabrier:2003} initial mass
function and solar metallicity, assuming exponentially declining
star-formation histories (SFHs) with $e$-folding times $\tau =
0.1$--10~Gyr. We allowed the galaxies to be attenuated with $0 \le A_{V} \le
6$~mag with  reddening following the \cite{Calzetti:2000} law. To
avoid strong influence on the fitting from one single band (in some
cases due to an emission line or bad photometry), we restricted the
maximum $S/N$ in the photometry to be 20. Examples of the fitting are
shown in Figure~\ref{fig:fast_sed_fit}.

For the few galaxies with spectroscopic redshifts in our sample,
Figure~\ref{fig:zspec_zphot} shows the comparison between
spectroscopic and photometric redshifts.  Our photometric redshift
estimates in general agree with spectroscopic redshifts with the
normalized median absolute deviation \citep{Brammer:2008}
$\sigma_{\rm NMAD} \sim 0.06$. (The significant outlier at $z_{\rm spec} =
1.2$ is a type-1 AGN.)\null\ However, the spectroscopic sample is
significantly biased with 9 of 11 objects detected in X-rays and a
median $H$-band magnitude $\langle H\rangle \approx 22.3$ (compared to
$\langle H\rangle \approx 23.9$ for
the total sample). Hence, larger and deeper spectroscopic samples are
needed to verify the accuracy of the photometric redshifts. On
the other hand, based on photometric redshifts estimates from CANDELS
\citep[see e.g., ][]{Dahlen:2013}, we derive $\sigma_{\rm NMAD} \sim
0.11$.  Specifically for HIEROs at $z \lesssim 3$, the
photometric redshifts from CANDELS are systematically higher by ${\sim}
0.2$--0.3, likely due to the absence of highly attenuated galaxy templates
in the CANDELS photometric redshift codes, (favoring a high redshift solution
to account for the red color actually due to attenuation). This trend
is also confirmed based on the photometric comparisons between ours
and from CANDELS for the whole HIERO sample.

\begin{figure*}[!htb] 
\centering
\includegraphics[trim= -10 -20 0 20,clip,width=0.48\linewidth]{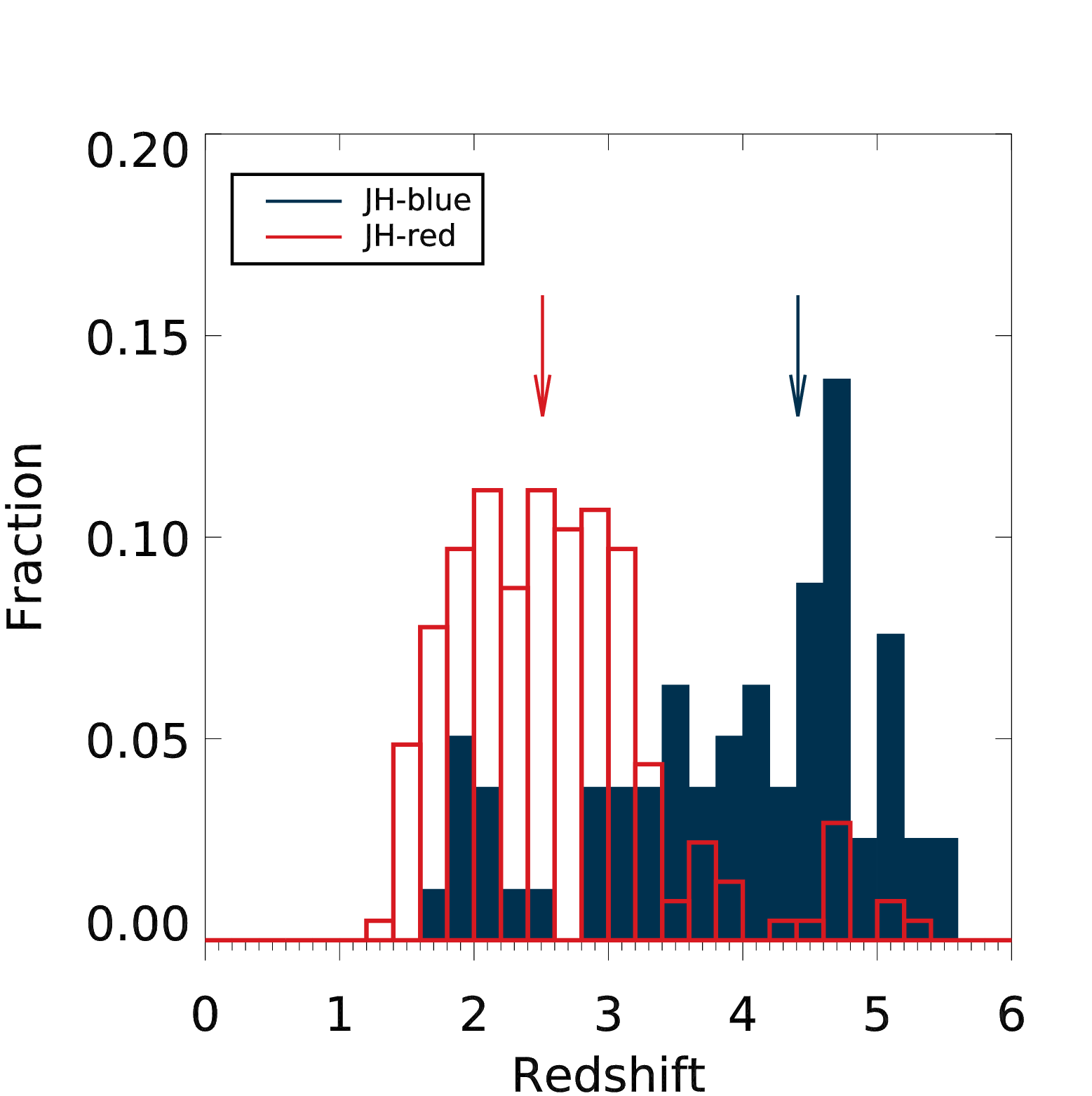}
\includegraphics[trim= 10 20 20 40,clip,width=0.5\linewidth]{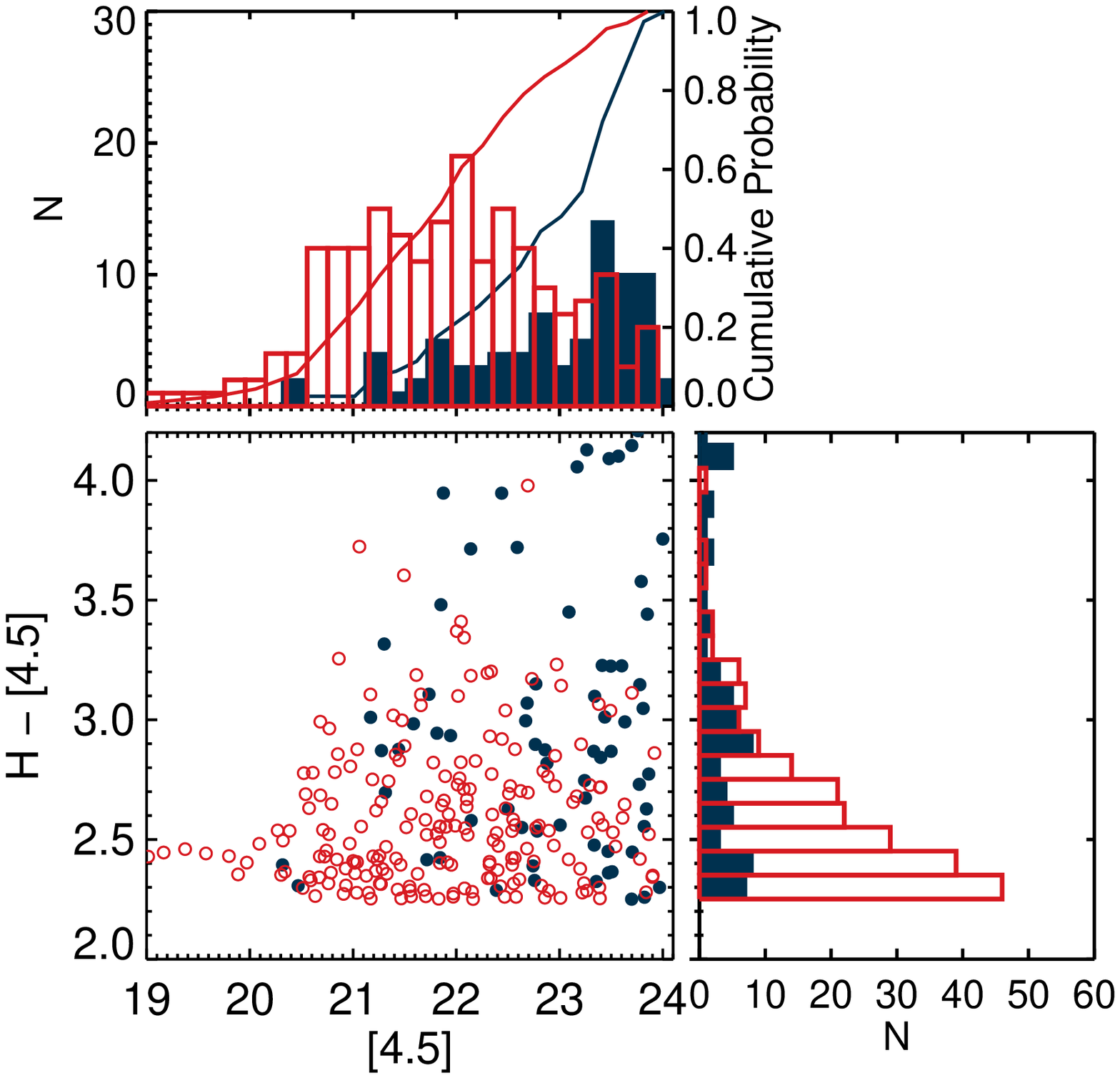}
\caption{\textbf{Left panel}: Photometric redshift distributions for
  the two categories of HEIROs. $JH\text{-}red$ and $JH\text{-}blue$
  HIEROs are shown by red empty and blue filled histograms,
  respectively, with their median values denoted by arrows.
  \textbf{Right panel}: $H - [4.5]$ versus [4.5] color-magnitude
  diagram for $JH\text{-}red$ (red) and $JH\text{-}blue$ (blue)
  HIEROs, as well as the histogram of $H - [4.5]$ and [4.5]
  magnitudes. The solid lines in the top plot represent the
  normalized cumulative distributions.\label{fig:zbest_hist}}
\end{figure*}

\begin{figure*}[!htb] 
\centering
\includegraphics[trim= -30 0 0 0,clip,width=0.49\linewidth]{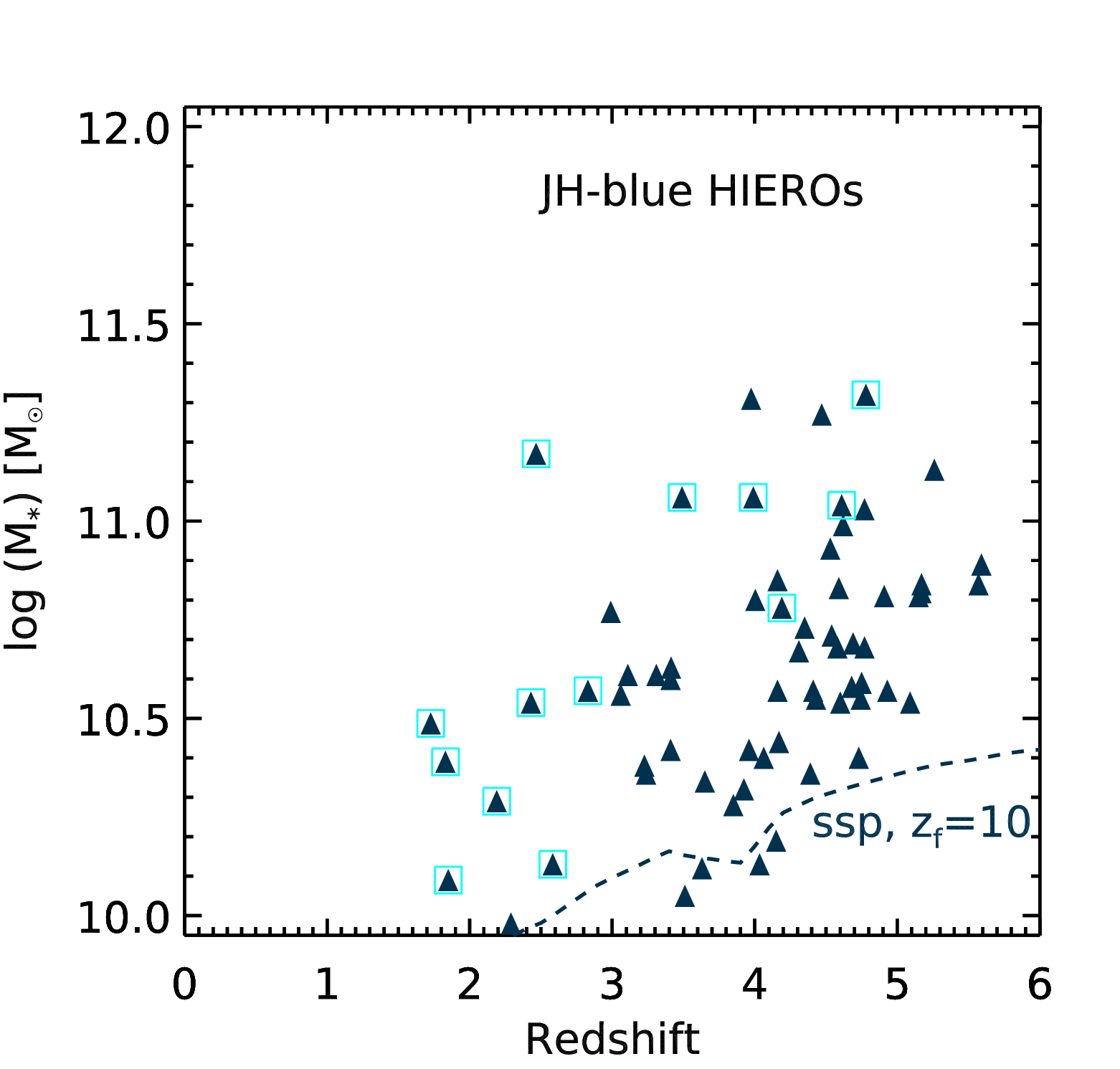}
\includegraphics[trim= -30 0 0 0,clip,width=0.49\linewidth]{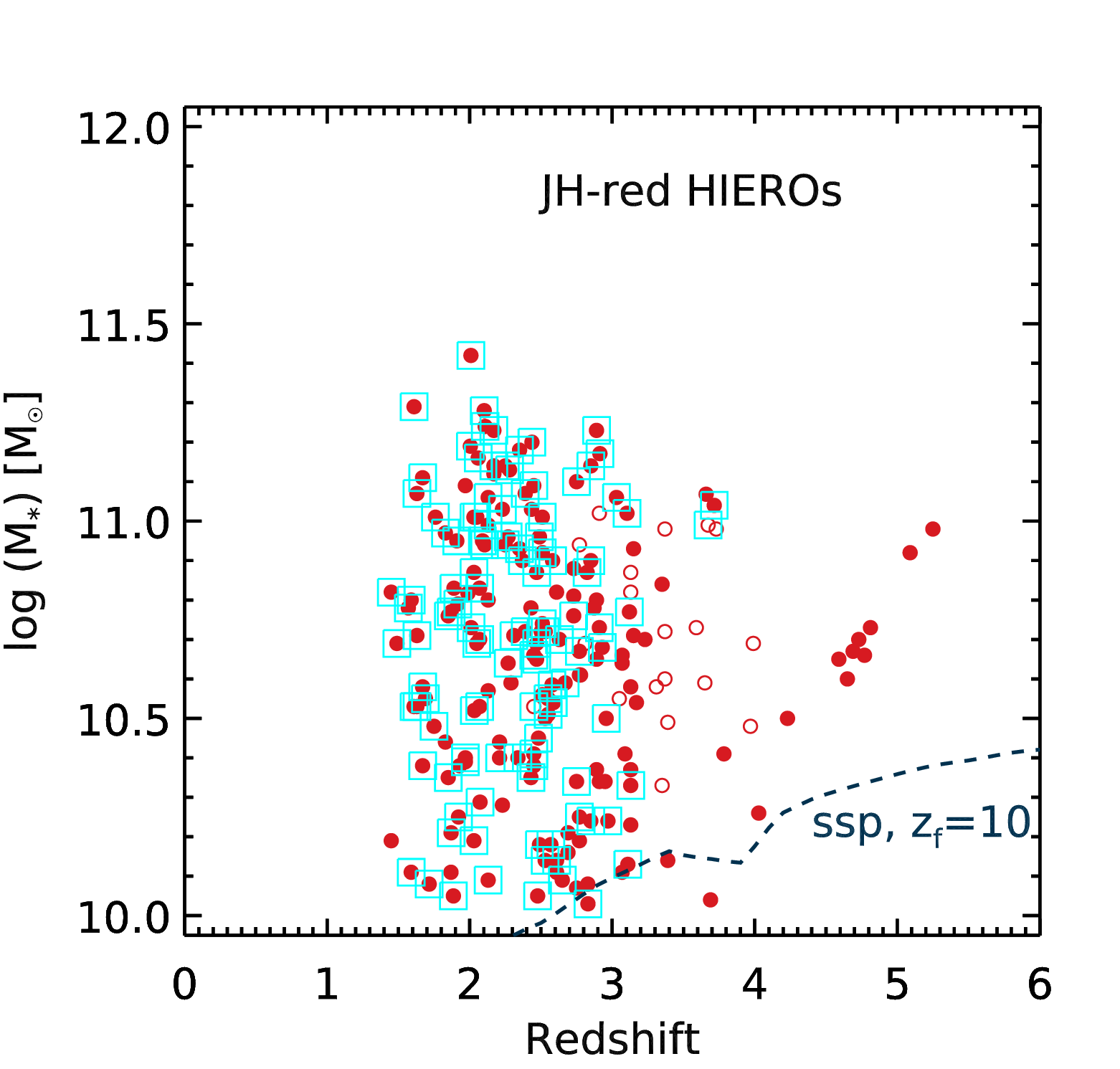}
\caption{Stellar mass versus redshift for $JH\text{-}blue$ HIEROs
  (left panel) and $JH\text{-}red$ HIEROs (right panel),
  respectively, based on the SED fitting with
  FAST\null. 24~$\mu$m-HIEROs are shown as cyan squares, which are 
  mostly at $z < 3$. In both panels, the blue dashed line
    denotes the mass completeness of our 4.5~$\mu$m selected sample
    ($[4.5] < 24$) as derived from an instantaneous-burst BC03 model
    formed at $z = 10$. In the right
    panel, galaxies that are classified as passive based on the
    color-color diagram in Figure~\ref{fig:rUVJ} are shown with red
    open circles.
\label{fig:mass_z}}
\end{figure*}

\subsection{Identifying sources with unreliable IRAC photometry}
\label{subset:clean}

While the majority of HIEROs yield a good fit, a substantial number
of galaxies can not be well fitted with FAST and the BC03
models. We checked in detail their SEDs and F160W and IRAC images and
found that most of them have bright
stars or galaxies within a few arcseconds, hence their
photometry is likely unreliable. This is particularly a problem in
the IRAC bands due to their larger PSFs. To explore whether this is
the origin of the problem, we re-fit the SEDs with the same
set of templates but excluding the IRAC photometry. A better fit
(much smaller $\chi^{2}$) was achieved for many sources with a
significantly different redshift
solution. Figure~\ref{fig:fast_zfit_bad} presents two such
examples. This illustrates that indeed the IRAC photometry for these
sources is problematic (in most cases boosted) due to contamination
by close neighbors. As a result, the true $H - [4.5]$ color
of these sources is likely much bluer (as can also be seen from their
best-fitted SED in the second fit).

Although in principle we can simply reject all sources with a bad fit
in the first run to clean our sample, it will introduce a bias
towards the choice of SED-fitting methods and templates, i.e., some
sources with a bad fit may be due to the wrong templates
instead of bad photometry. To be conservative, we borrow the ``clean
index" concept in dealing with the source confusion at far-infrared
wavelengths \citep{Elbaz:2011} to identify sources whose IRAC fluxes
are reliable. We define a source as ``clean" only if it satisfies at
least one of the following two criteria: (a) it has no neighbor
within 2$\arcsec$ (rougly the size of the FWHM of the IRAC PSF) and a
good SED(photometric redshift)-fitting result ($\chi^{2} < 4$,
corresponding roughly to the one-tailed (right-tail)
probability $>$ 0.95) when including IRAC photometry; (b) similar
redshift solutions are achieved during the two SED-fitting runs,
i.e., $| z_{\rm phot}({no~IRAC})-z_{\rm phot}({IRAC})|/z_{\rm phot}({IRAC}) < 30\%$ (For
sources that only satisfy the second criterion, it is unclear whether
their IRAC photometry is contaminated or not: the bad fit could 
be because there are no perfect templates in the library or because
the data quality is not good enough.) This leaves us 285 clean HIEROs
including the 18 $H$-dropouts (Table~\ref{tab:number_counts}). Among
these 285 HIEROs, 223 of them (78\%) have a good fit with $\chi^{2}
< 4$ when including the IRAC photometry.

Figure~\ref{fig:zbest_hist} shows the photometric redshift
distribution for the final clean samples. Our classification based on
$J-H$ and $H - [4.5]$ colors successfully separates galaxies at $z
\gtrsim 3.5$ from those at relatively lower redshifts, i.e., $2 < z <
3$. The median redshifts for $JH\text{-}blue$ and $JH\text{-}red$
HIEROs are $\langle z\rangle \sim 4.4$ and $\langle z\rangle \sim 2.5$, respectively. Among
the 79 $JH\text{-}blue$ HIEROs, only 14 (18\%) have photometric
redshifts $z_{\rm phot} < 3$. Most of these 14
are detected at MIPS 24~$\mu$m, in contrast to the $z > 3$
$JH\text{-}blue$ HIEROs. On the other hand, 52 out of 206 (28\%) of
$JH\text{-}red$ HIEROs are at $z > 3$ with the majority (30 out of
52) being at $3 < z < 3.5$. Figure~\ref{fig:zbest_hist} also
compares the distribution of the two populations of HIEROs in the $H
- [4.5]$ versus $[4.5]$ color-magnitude diagram. As expected, the
$JH\text{-}blue$ HIEROs are in general fainter at 4.5~$\mu$m
as well as redder in  $H - [4.5]$.

\begin{figure}[!htb]
\centering
\includegraphics[trim= -30 -40 0 0,clip,width=0.95\linewidth]{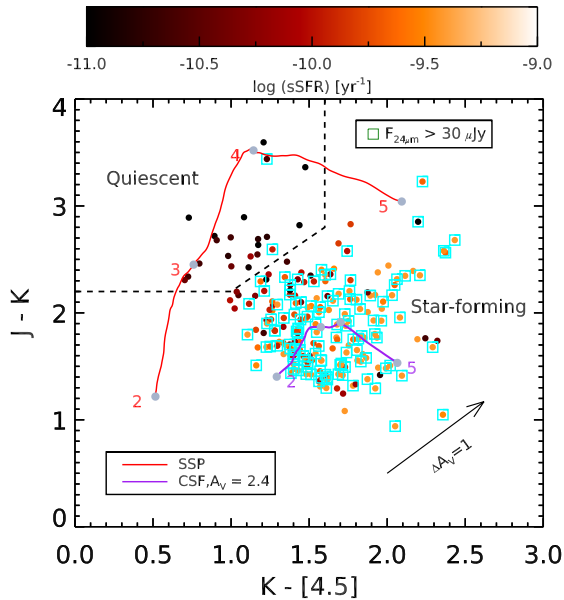}
\caption{Observed $J - K$ versus $K - [4.5]$ diagram for $JH\text{-}red$ HIEROs in our sample. 
Solid lines show the $2 < z < 5$  color tracks for a passive galaxy with an
age of 1.0~Gyr and a star-forming galaxy with an age of 0.3~Gyr and
constant star formation history. All galaxies are color-coded with
their best-fit specific star formation rates (sSFR). The dashed line
denotes the dividing line that separates quiescent from star-forming
galaxies. The arrow shows the effect of 1~mag of dust extinction at
$z=3$ {\bf assuming a Calzetti law}. 
\label{fig:rUVJ}}
\end{figure}

\section{Stellar populations of HIEROs}
\label{Sec:classification}
\subsection{$JH\text{-}blue$ HIEROs: Normal massive star-forming
  galaxies at $z > 3$} 

Our analysis on photometric redshifts (both from our own estimates
based on FAST and those from CANDELS) confirms that the selection
criterion of $JH\text{-}blue$ HIEROs yields a clean selection of $z >
3$ galaxies. Remaining low-redshift contaminants can be removed
based on SED modeling or on their strong 24~$\mu$m detections (with a
median $F_{24~\micron} \sim 98$~\uJy). The F160W images of these $z <
2$ contaminants show that most of them are either extremely faint or
likely mergers, leading to uncertain $J - H$ colors or unusual
extinction properties, which explains why they enter our selection.

The advantage of using FAST to derive photometric redshifts is that
it also gives self-consistent estimates of stellar masses. 
Figure~\ref{fig:mass_z} shows the stellar mass estimates
of $JH\text{-}blue$ HIEROs. The modeling
assumptions are described in Section~\ref{Sec:redshifts}, i.e., the
BC03 stellar library, \cite{Calzetti:2000} extinction law with $0 \le A_{V}\le
6$, and exponentially declining star formation histories (single
$\tau$ model) with $e$-folding times ranging from 0.1 to 10
Gyr. Delayed $\tau$ models give nearly identical estimates of stellar
masses.  Our mass estimates imply that the $JH\text{-}blue$ HIEROs
are massive star-forming galaxies at $z \gtrsim 3.5$ with a median
$\langle M_{*}\rangle \approx 10^{10.6}$~\Msol.

The best-fit ages for $JH\text{-}blue$ HIEROs from single $\tau$
models range from $t = 0.1$ to $t$ = 1.6 Gyr with a median 
$\langle t\rangle\approx 1$~Gyr for galaxies at $z > 3$ and 
$\langle t\rangle\approx 0.1$~Gyr for galaxies a $z
\leq 3$. The median attenuation is $\langle E(B-V)\rangle = 0.33$,
with galaxies at $z < 3$ much dustier than those at $z > 3$ 
($\langle E(B-V)\rangle = 0.83$ vs.\ $\langle E(B-V)\rangle = 0.25$). These best-fit stellar
properties do not change significantly by using delayed $\tau$
models. These results suggest that most $JH\text{-}blue$ HIEROs are
massive, dusty star-forming galaxies which have already assembled
relatively old stellar populations. Their red $H - [4.5]$ colors thus
appear to be caused by a combination of moderately old stellar
populations (strong 4000~\AA~break) and dust attenuation. The few $z
< 3$ contaminants tend to be less massive, younger, and also much
dustier, suggesting that the red $H - [4.5]$ colors are mostly caused
by severe dust attenuation.

\subsection{$JH\text{-}red$ HIEROs: Massive dusty star-forming
  galaxies at $2 < z < 3$ and passive galaxies at $3 < z < 4$} 

As discussed in Section~\ref{subsec:sample} and
Section~\ref{Sec:redshifts}, $JH\text{-}red$ HIEROs include in
general two populations, $z \sim 2$--3 dusty star-forming galaxies and
passive galaxies at $z \sim 3$--4. As illustrated in
Figure~\ref{fig:sed_comp}, it is possible to distinguish between the
two populations based on their different behaviors in the observed
near-infrared to mid-infrared colors. Here we propose to separate the
two populations with the observed $J - K$ versus $K - [4.5]$
color-color diagram, which for $z \sim 3$ resembles the rest-frame $U - V$ versus $V
- J$ diagram~\citep{Williams:2009}. 
Figure~\ref{fig:rUVJ} shows the distribution of
$JH\text{-}red$ HIEROs in this
diagram. Galaxies with lower sSFR (from SED fitting) populate
the upper left region while galaxies with higher sSFR locate in the
lower right region. As independent evidence, most
24~$\mu$m detections fall in the star-forming region, consistent with
star-forming galaxies at $z < 3$.
Based on the distribution of galaxies with different sSFRs and the
direction of reddening from the \cite{Calzetti:2000} law, we define
the criteria for quiescent galaxies as : $J - K > (K - [4.5]) + 1.2$,
$J - K > 2.2$, and $K - [4.5] < 1.6$. The remaining galaxies are
classified as star-forming. This diagram is not only limited to classifications of
HIEROs but also provides  an efficient way to identify quiescent
galaxies at $z \gtrsim 3$ in general.

As shown in Figure~\ref{fig:rUVJ}, roughly 10\% (21 out of 206) of
the $JH\text{-}red$ HIEROs are classified as quiescent
galaxies. The median redshift of these is $\langle z\rangle\approx 3.4$,
suggesting that they are among the earliest quenched systems in the
Universe. Among these 21 quiescent $JH\text{-}red$ HIEROs, 15
are at $z_{\rm phot} > 3$ out  of 52 $z >
3$ $JH\text{-}red$ HIEROs altogether. 

The $JH\text{-}red$ HIEROs are also primarily massive galaxies (the
right panel of Figure~\ref{fig:mass_z}) with a  $\langle M_{*}\rangle\approx
10^{10.7}$~\Msol\ for both quiescent and star-forming
subpopulations. The median best-fit age and attenuation for quiescent
galaxies are $\langle t\rangle = 1.0$~Gyr and $\langle E(B - V)\rangle = 0.17$ while for
star-forming galaxies $\langle t\rangle = 0.7$~Gyr and $\langle E(B - V)\rangle = 0.6$,
respectively. These results are consistent with the classifications
based on the observed $J - K$ versus $K - [4.5]$ diagram. The high
attenuation value for star-forming galaxies in this $JH\text{-}red$
HIERO population is also consistent with their high 24~$\mu$m flux
densities with  $\langle F_{24~\micron}\rangle\approx 90~$\uJy, which
corresponds roughly to a total infrared luminosity $L_{\rm TIR} \sim
10^{12}$~\Lsol\ at $z \sim 2.5$.

Based on X-ray luminosity estimates, ${\sim}20\%$ of $JH\text{-}red$
HIEROs are classified as X-ray AGNs ($L_{\rm 0.5-8~keV} > 10^{42}$~erg~s$^{-1}$),
consistent with the X-ray AGN fraction in massive galaxies
at $z\sim 2$ (Wang et al. 2015, submitted). The $JH\text{-}blue$ HIEROs 
have similar AGN fraction,
${\sim}18\%$. These results indicate that our selection criteria are
not particularly biased towards AGNs compared to non-AGN massive
galaxies.

\begin{figure}[htb] 
\centering
\includegraphics[trim= -20 -30 20 0,clip,width=0.95\linewidth]{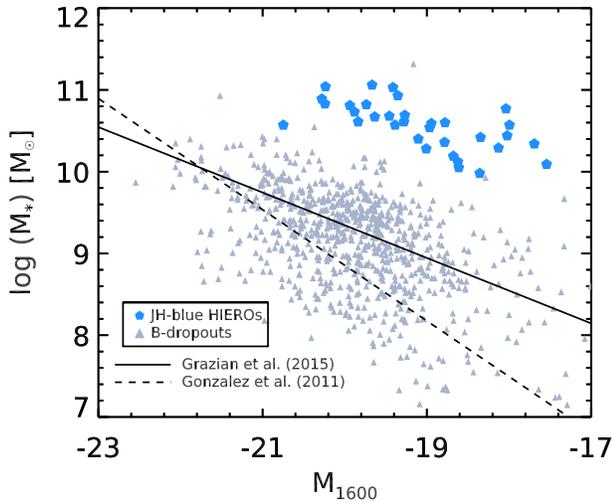}
\caption{Stellar masses as a function of UV luminosity for LBGs
  ($B$-dropouts, triangles) and $JH\text{-}blue$ HIEROs (pentagons) in GOODS fields. Only $H$-detected HIEROs 
  are shown to ensure that the rest-frame UV magnitudes are reliable. The
  best-fit $M_{*}$--$L_{\rm UV,1600}$ relation for
  $B$-dropouts \citep{Gonzalez:2011} and mass-selected
  samples \citep{Grazian:2015} at $z\sim4$ are shown with the dashed
  and solid line, respectively.   
  \label{fig:lmass_UV}}
\end{figure}

\begin{table*}\centering
\ra{1.3}
\caption{Stacked infrared flux densities of HIEROs\label{tab:stacked_flux}}
\begin{tabular}{@{}lcccccccccc@{}}
\toprule
Sample & 16~$\mu$m & 24~$\mu$m & 70~$\mu$m & 100~$\mu$m & 160~$\mu$m & 250~$\mu$m & 350~$\mu$m & 500~$\mu$m & 870~$\mu$m &  1.1 mm \\
& [$\mu$Jy] & [$\mu$Jy] & [$\mu$Jy] & [mJy] & [mJy]& [mJy]& [mJy]& [mJy]& [mJy]& [mJy] \\
\hline
$JH\text{-}blue$ HIEROs & 5.8$\pm$3.7 & 14.6$\pm$3.8 & 184$\pm$260 & 0.22$\pm$0.05 & 0.90$\pm$0.20 & 3.09$\pm$0.54 & 3.67$\pm$0.83 & 4.60$\pm$0.62 & 1.15$\pm$0.19 & 0.92$\pm$0.22 \\
$JH\text{-}red$ HIEROs(SF) & 28.6$\pm$5.9 & 74.4$\pm$7.4 & 177.7$\pm$27.2 & 0.63$\pm$0.04 & 2.01$\pm$0.16 & 6.21$\pm$0.55 & 7.22$\pm$0.64 & 5.61$\pm$0.49 & 1.09$\pm$0.14 & 0.72$\pm$0.15 \\
$JH\text{-}red$ HIEROs(QS) &  5.7$\pm$5.6 & 6.3$\pm$3.4 & 84.3$\pm$40 & 0.1$\pm$0.05 & 0.2$\pm$0.1 &0.3$\pm$0.3 &0.3$\pm$0.5 &2$\pm$0.7 &-0.06$\pm$0.1 &-0.4$\pm$0.1 \\  
\bottomrule
\end{tabular}
\end{table*}

\begin{figure*}[!htb] 
\centering
\includegraphics[trim= -20 0 0  0,clip,angle=0,width=0.48\linewidth]{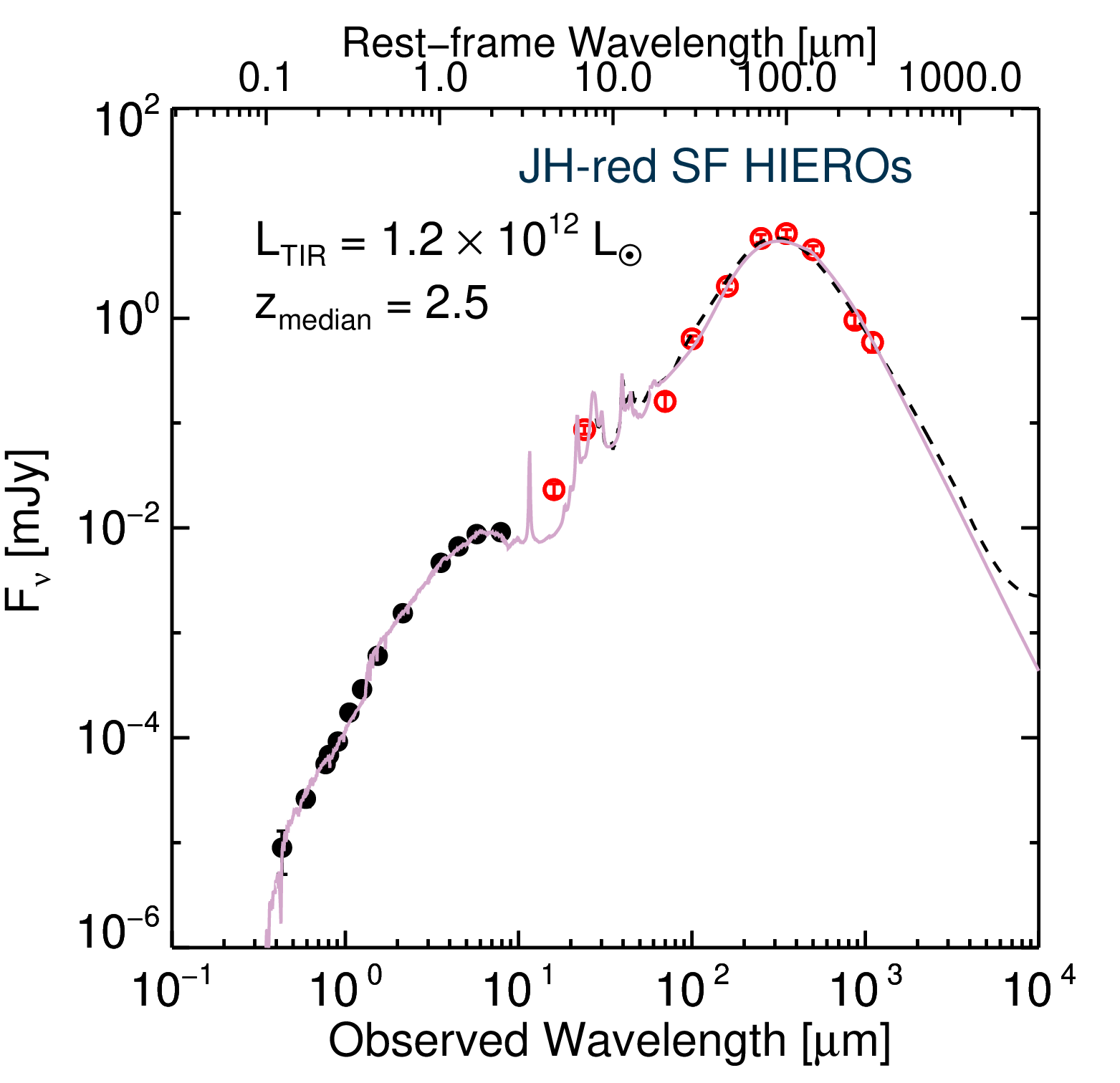}
\includegraphics[trim= -20 0 0  0,clip,angle=0,width=0.48\linewidth]{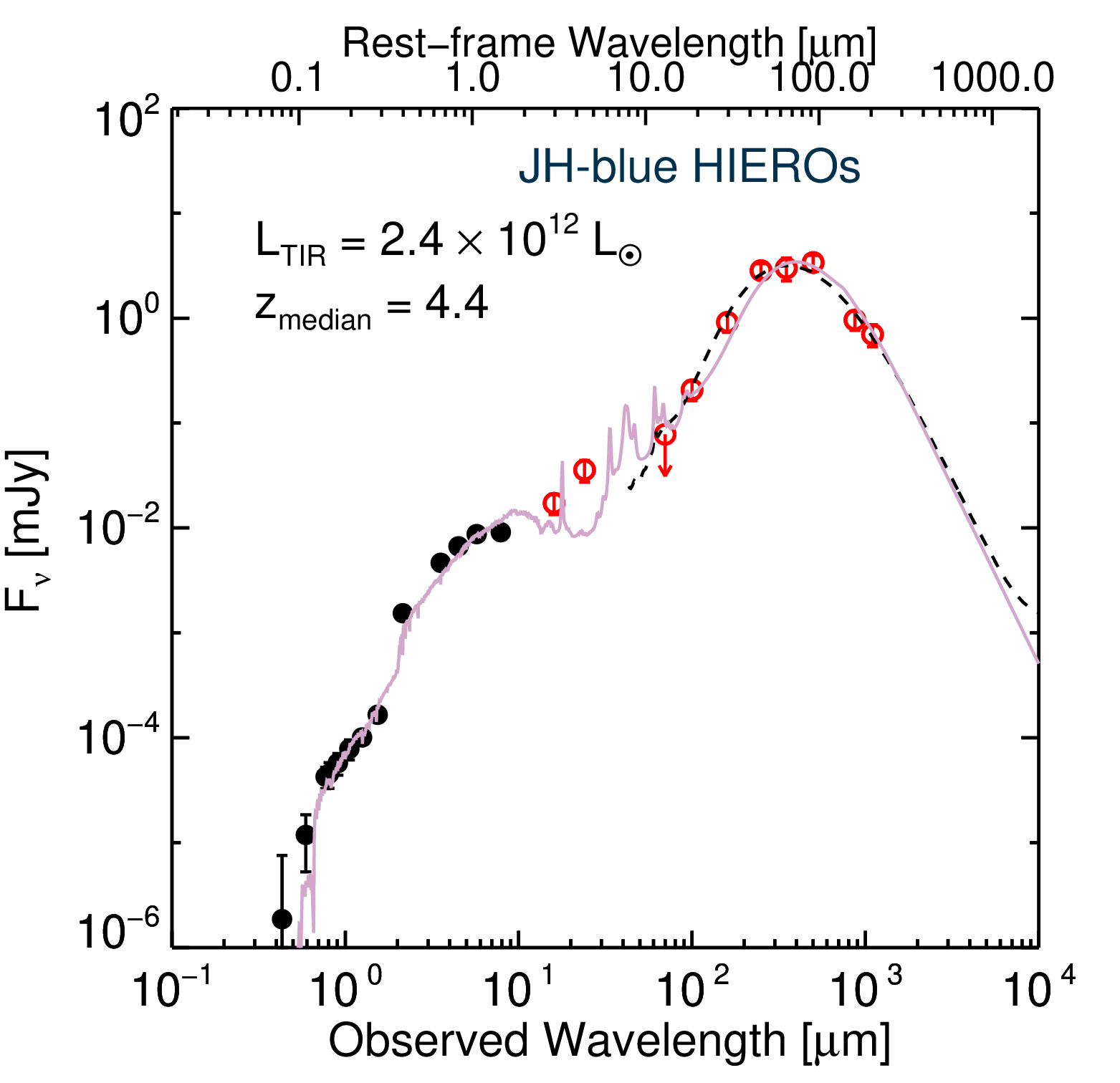}
\caption{Stacked SEDs for $JH\text{-}red$ SF (left) and
  $JH\text{-}blue$ (right) HIEROs. Median flux densities between
  F435W and F160W bands were derived based on stacking of the
  corresponding {\it HST} images, while the median $K$-band and IRAC flux
  densities were derived from the measured fluxes of individual
  sources. The best-fitting IR SED templates from \cite{Chary:2001}
  are shown as dashed curves, while the best-fitting full-band SEDs
  from CIGALE are shown with solid lines. The derived total infrared
  luminosities and median redshifts for the two populations are also
  shown.  
  \label{fig:sed_stacking_hz}} 
\end{figure*}

\section{Star formation properties of  HIEROs}
\label{Sec:SF}
\subsection{UV properties}

By selection, the HIEROs are faint in their rest-frame UV despite
their high stellar masses. Specifically, the $JH\text{-}blue$ HIEROs
are generally massive star-forming galaxies at $z > 3$. Previous
studies of LBGs at these redshifts reveal a tight correlation between
stellar mass and UV luminosity~\citep[see, e.g.,
][]{Gonzalez:2011}. This allows the
UV luminosity function to translate directly to the stellar mass function, as is
commonly adopted in studying high-redshift stellar mass functions. It
is therefore interesting to compare HIERO and LBG stellar masses at
similar redshifts and UV luminosities to  explore how the HIEROs
affect the stellar mass estimate.

We selected a sample of $z \sim 4$ LBGs ($B$-dropouts) that are on
average at similar redshifts as the $JH\text{-}blue$ HIERO
population in the two GOODS fields using the same criteria as
\cite{Bouwens:2012}. The rest-frame UV luminosities ($M_{1600}$) were
derived using $EaZY$ \citep{Brammer:2008} at fixed redshifts from
CANDELS\null.  Figure~\ref{fig:lmass_UV} compares the
$M_{*}$--$L_{\rm UV}$ relations.  (The HIERO rest-frame UV
luminosities came from the best-fit SED templates at fixed redshifts
estimated by FAST\null.)  Compared to LBGs at similar masses, the HIEROs are
generally 2--3~magnitudes fainter in the rest-frame UV\null. The
existence of these galaxies suggests that using a simple
$M_{*}$--$L_{\rm UV}$ relation (or using only UV-selected samples) to
determine the stellar mass function underestimates the massive
end. We will illustrate this point further in
Section~\ref{Sec:discussion}.

\begin{table*}\centering
\ra{1.3}
\caption{Measured physical properties of HIEROs\label{tab:statistics_sfr}}
\begin{tabular}{@{}cccccccc@{}}
\toprule
Number & $n$ & $z_{\rm phot}$ & log $M_{*}$ & log ($L_{IR}$)  & UV slope ($\beta$) & SFR & sSFR \\ 
& [arcmin$^{-2}$] & [median] & [mean, \Msol] &[\Lsol] & &  [\Msol\ yr$^{-1}$] &[Gyr$^{-1}$]\\
\hline
$JH\text{-}blue$ HIEROs \\

66      & 0.2  & 4.43 & 10.78  & 12.38 & -0.05 & 240 & 4.2$_{-0.8}^{+0.6}$ \\ 
$JH\text{-}red$ HIEROs (SF)\\
185 & 0.54 & 2.51 & 10.77   & 12.10 & 1.12 & 120 & 1.9$_{-0.2}^{+0.2}$\\ 
$JH\text{-}red$ HIEROs (QS)\\
21 & 0.06 & 3.42 & 10.7 & $< 11.65$ & -- & $< 45$ & $ < 0.9$ \\
\bottomrule
\end{tabular}
\end{table*}

\subsection{Determining total infrared luminosities of HIEROs through stacking }
\label{Sec:stacking} 

The general faintness of HIEROs in the UV suggests that most of their
star formation is hidden by dust. The amount of this hidden star
formation can be inferred through infrared emission, which originates
from the thermal reradiation by dust of their absorbed ultraviolet
light. Therefore, understanding the Spectral Energy Distributions
(SEDs) in the infrared is essential to get a complete view of star
formation in HIEROs. Alternatively, we could  use dust-unbiased
tracers of star formation, e.g., radio continuum, to measure the
total star formation rate. However, although both the far-infrared
and radio surveys in the GOODS fields are among the deepest ever
conducted, only a few HIEROs are individually detected. Moreover,
most of the detected ones are $JH\text{-}red$ HIEROs. For instance,
crossmatching with the VLA 1.4~GHz catalog in GOODS-North
\citep{Morrison:2010}, 33 out of 142 HIEROs are detected at 1.4~GHz
with $F_{\rm 1.4~GHz} > 20$~uJy, but only 6 out of 33  are
classified as $JH\text{-}blue$ HIEROs. This is likely caused by the
high redshifts of the $JH\text{-}blue$ HIERO population and
that most of them are not extreme starbursts (contrary to bright
submillimeter galaxies).  To probe lower star formation rates that
are typical of HIEROs, a stacking approach is required.

For the $JH\text{-}blue$ HIERO population, we excluded contaminants at
low redshifts by applying a redshift cut of $z > 3$, which leaves us
66 galaxies. For the $JH\text{-}red$ HIERO population, we separately
stacked star-forming (SF) and quiescent (QS) galaxies.  The
exquisite multi-wavelength data in GOODS allow us to perform stacking
across the whole infrared wavelength range, including the 16 and
24~$\um$ bands from {\it Spitzer}, 100, 160, 250, 350 and 500~$\um$ from
{\it Herschel}, 850~$\mu$m from SCUBA, 870~$\um$ from LABOCA, and 1.1~mm
from AzTEC. This permits a comprehensive understanding of their
infrared SEDs. Moreover, the combination of stacked far-infrared and
submillimeter colors provides an independent and complementary
estimate of their redshift from the position of their peak
far-infrared emission~\citep{Hughes:2002,Daddi:2009a}.

To avoid contamination from a few relatively bright members, we
conducted a median stacking in the infrared and submillimeter bands
(16~\micron--1.1~mm) using the IAS stacking code
library~\citep{Bethermin:2010}. (Using 
mean stacking would leave the main results unchanged, likely because
the fraction of bright members is small.)  We retrieved 
flux densities by PSF-fitting  the stacked images. A
correction factor ranging from 0.92 at 250 $\micron$ to 0.75 at
500~$\micron$ was applied to account for clustering, which does not
change much with redshift and stellar mass of the
galaxies~\citep{Bethermin:2012,Schreiber:2015}. We determine
uncertainties on the flux densities with bootstrapping.

The stacked flux densities for all populations are listed in
Table~\ref{tab:stacked_flux}. For $JH\text{-}blue$ HIEROs,
significant detections are revealed except at 16 and 70~\micron\ due
to shallower depths. For $JH\text{-}red$ SF HIEROs, significant
detections are found at all wavelengths while there are no
detections with $S/N > 3$  in any infrared bands for
$JH\text{-}red$ QS HIEROs. This provides independent evidence that
our approach successfully separates the QS and SF
populations. Moreover, the peak of the infrared SED for
$JH\text{-}blue$ and $JH\text{-}red$ SF HIEROs falls respectively at
${\sim} 500$~\micron\ and $\sim$350~\micron, lending evidence that they are
most likely at $z \sim 4$ and $z \sim 2.5$, consistent with our
photometric redshifts.
\begin{figure*}[!htb] 
\centering
\includegraphics[trim= 0 0 0 0,clip,width=0.49\linewidth]{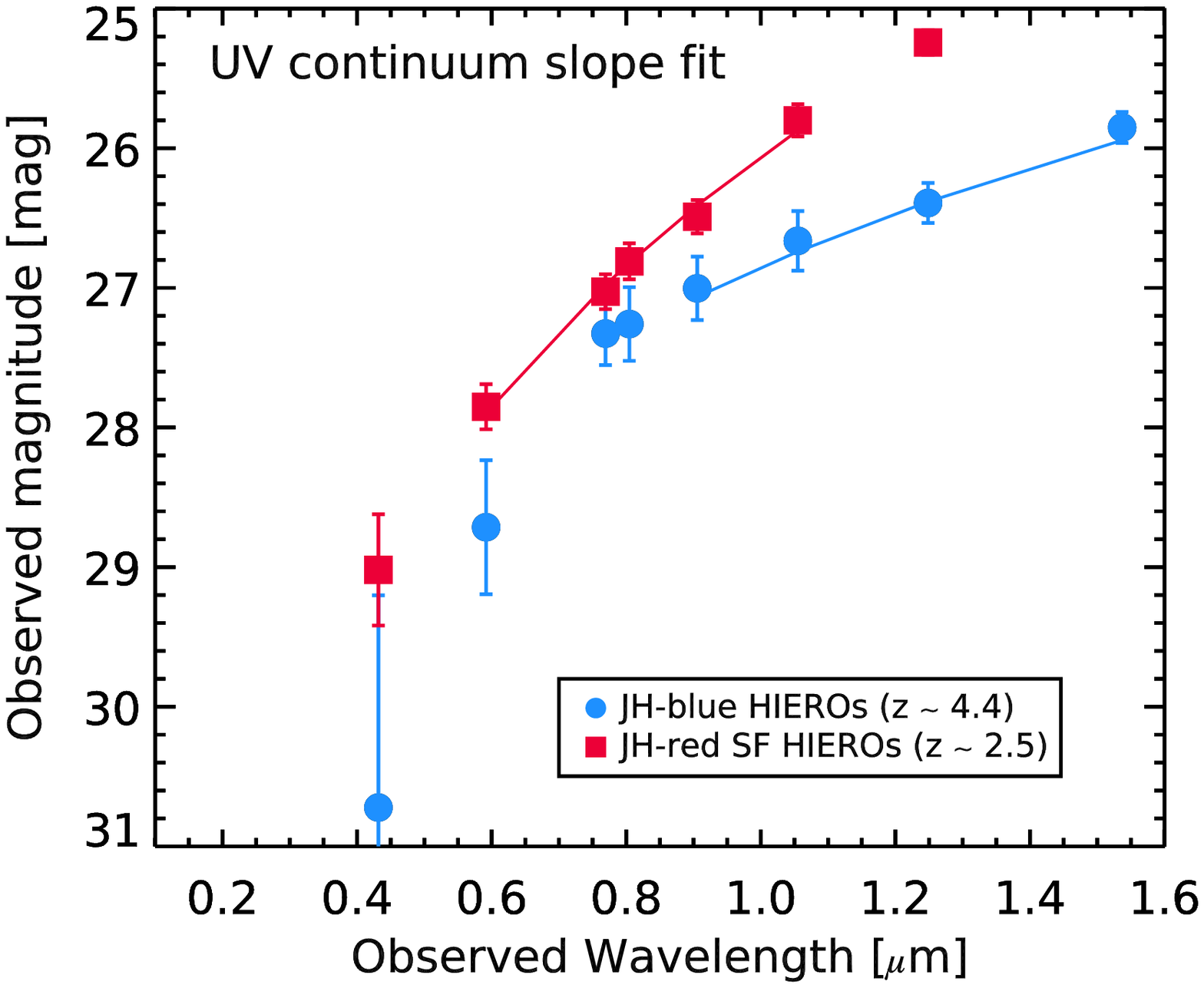}
\includegraphics[trim= 0 0 0 0,clip,width=0.49\linewidth]{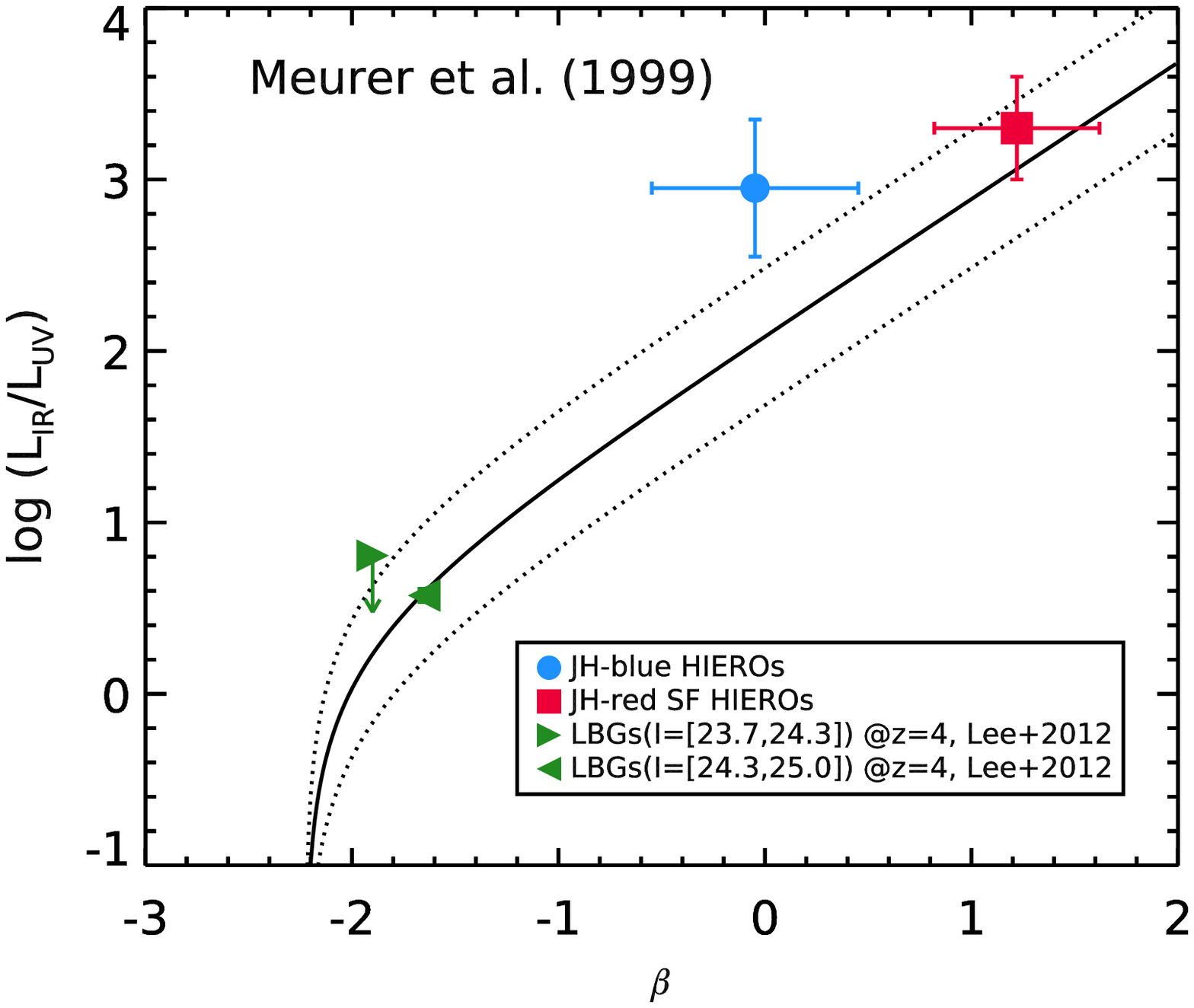}
\caption{\textbf{Left panel}: Illustration of how we estimated the UV
  continuum slope for $JH\text{-}red$ SF (red) and
  $JH\text{-}blue$ (blue) HIEROs.  The strong break
  between observed-frame $B_{435}$, $V_{606}$, and $I_{775}$
  bands for $JH\text{-}blue$ HIEROs provides independent evidence that
  a significant fraction of them should be at similar redshifts as
  $B_{435}$- and $V_{606}$-dropouts, i.e., $z \sim
  4$--5. \textbf{Right panel}: IRX ($L_{IR}/L_{UV}$) values vs.\ UV
  slope ($\beta$) for HIEROs and LBGs. The \cite{Meurer:1999}
  relation is shown with 0.4~dex scatter.
\label{fig:extinction_correction}}
\end{figure*}

\begin{figure*}[!htb] 
\centering
\includegraphics[trim= -30 0 0 0,clip,angle=0,width=0.48\linewidth]{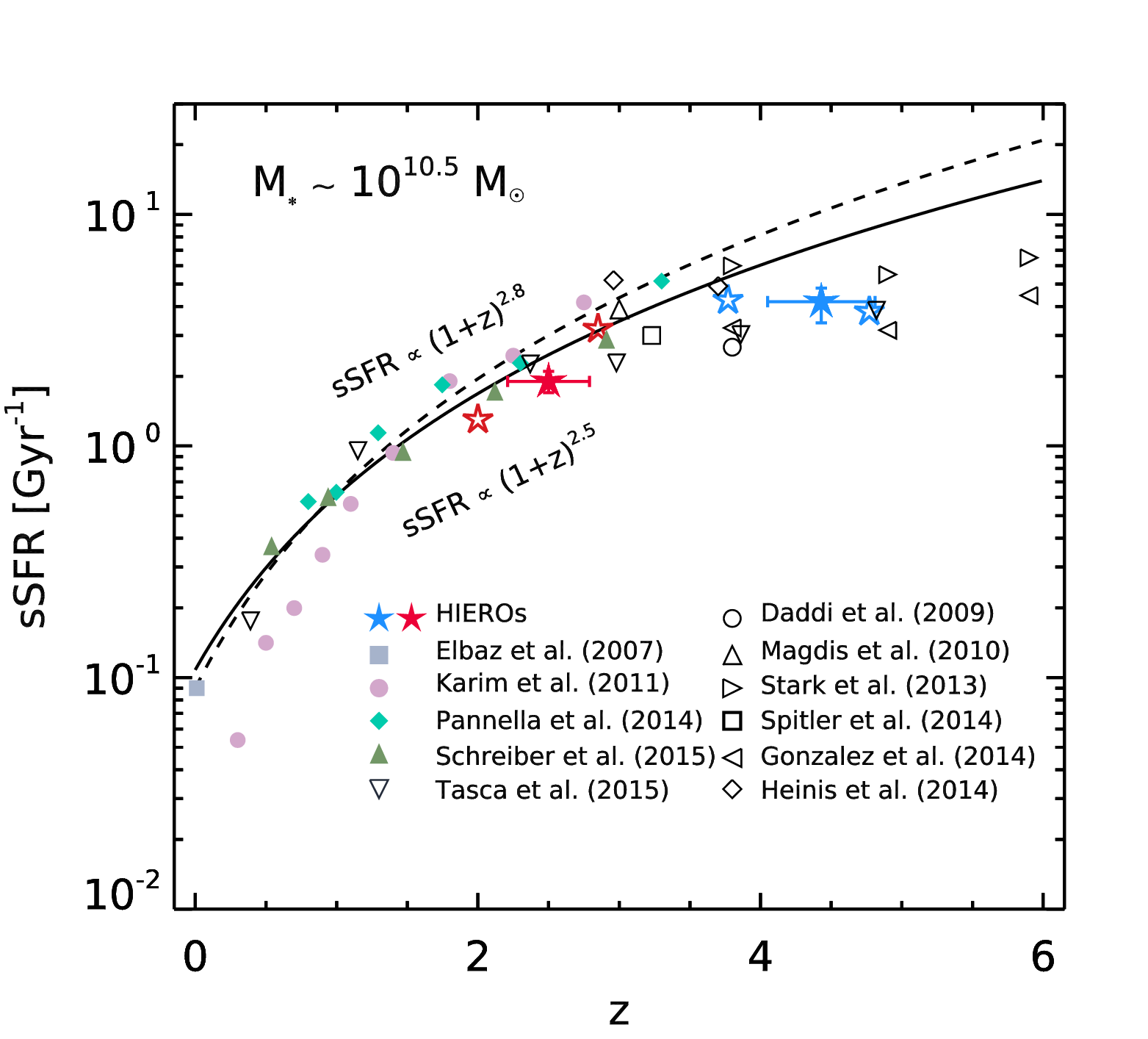}
\includegraphics[trim= -30 0 0 0,clip,angle=0,width=0.48\linewidth]{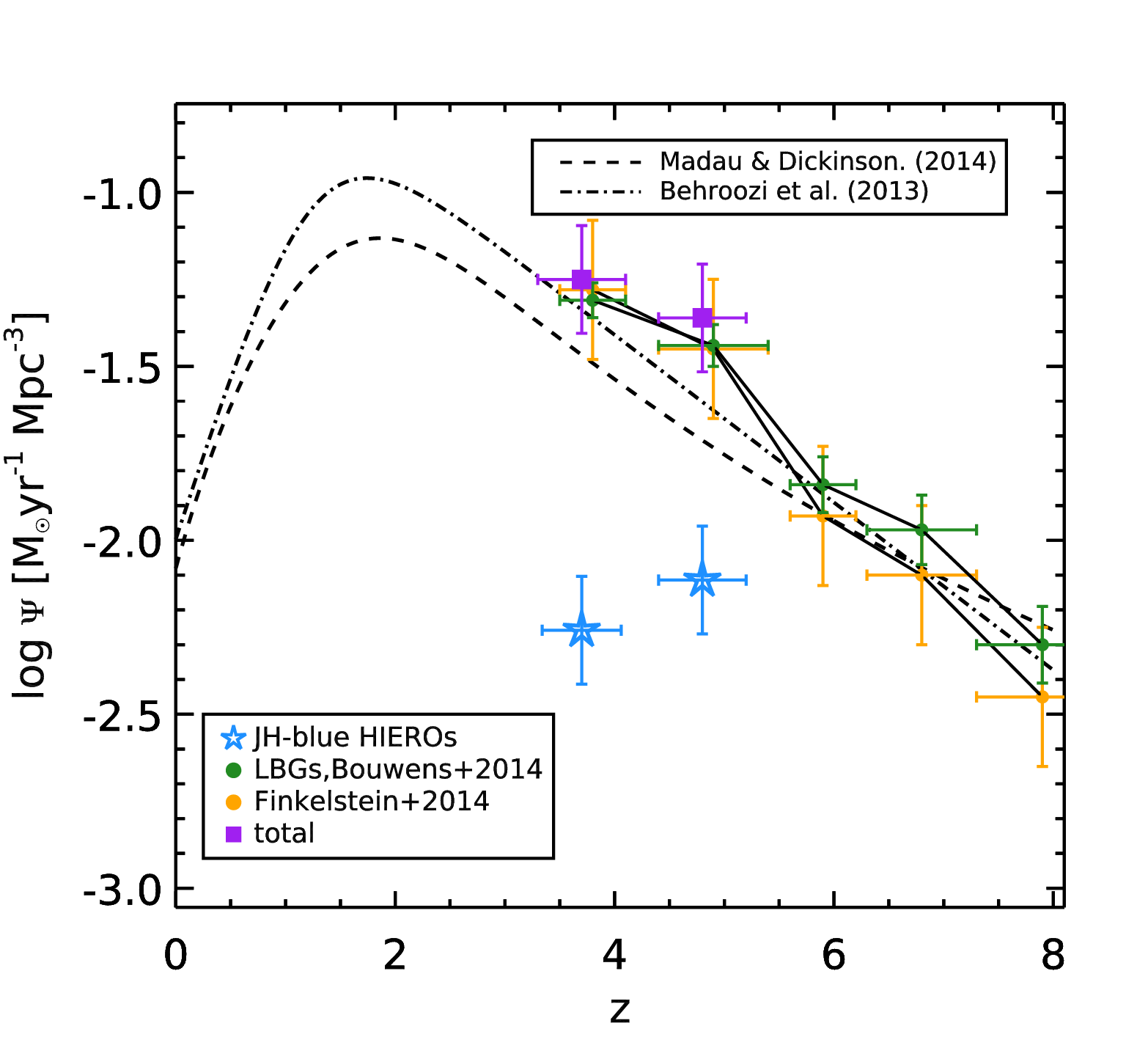}
\caption{\textbf{Left panel}: Specific SFR of star-forming galaxies
  with stellar mass $M_{*} \sim 10^{10.5}$~\Msol\ as a function of
  redshift. The sSFRs of $JH\text{-}blue$ ($z > 3$) and
  $JH\text{-}red$ SF HIEROs are shown as red and blue filled
  pentagrams, respectively. The horizontal error bars
    represent the 1$\sigma$ scatter of the redshift distribution. We
    also derived sSFR for each population of HIEROs in narrower
    redshift bins, which are shown with red and blue open
    pentagrams for $JH\text{-}red$ and $JH\text{-}blue$ HIEROs,
    respectively. The dashed line indicates the best-fit of the
  sSFR evolution at $z < 2$  \citep{Sargent:2014}, while the solid
  line indicates the predicted sSFR evolution from theoretical models
  \citep{Dekel:2014},  normalized at $z \sim
  1$. Given that the HIEROs represent the majority of
    galaxies with $M_{*} > 10^{10.5}$~\Msol, our results suggest
    that the sSFR for massive galaxies continues to increase
    up to $z \sim 4$ but then becomes relatively flat at higher
    redshifts. \textbf{Right panel}: The evolution of star formation
  rate density (SFRD) as a function of redshift. The dashed and
  dotted lines denote the evolution of SFRD presented by
  \cite{Madau:2014} and \cite{Behroozi:2013a},
  respectively. The blue open pentagrams show the SFRD
    of $JH\text{-}blue$ HIEROs at two redshift bins: $3 < z < 4.4$
    and $4.4 < z < 6$, the same redshift bins as shown by
    the blue open pentagrams in the left panel. Compared to previous
    SFRD measurements based on LBGs at these redshifts, the HIEROs
    make up $15-25\%$ of that by LBGs.
  \label{fig:ssfr_z_massive}}
\end{figure*}



Figure~\ref{fig:sed_stacking_hz} shows the median SEDs of the two
populations of HIEROs.  With the well-constrained SED shape, 
their total infrared luminosities (TIR)  constrain
their star formation rates. We fit the stacked 160~$\mu$m$-$1.1 mm
(to avoid rest-frame ${<} 40$~\micron\ wavelengths, which may suffer from
AGN contamination) SEDs using a suite of infrared templates,
including the 105 template SEDs from \cite{Chary:2001} (CE01,
hereafter). During the fitting, we fixed the redshift at the median
value of the sample and left the template normalizations as free
parameters. The best-fit model is the template that minimizes
$\chi^{2}$.  The total infrared luminosity for $JH\text{-}red$ and
$JH\text{-}blue$ HIEROs are $1.2 \times 10^{12} L_{\odot}$ and $2.4
\times 10^{12} L_{\odot}$, respectively. We also fit the full-band
SED from UV to far-IR for HIEROs using the code CIGALE
\citep{Noll:2009,Serra:2011} using BC03 models in the UV-to-NIR and
\cite{Draine:2007} models in the infrared. This gives a consistent
measurement of $L_{TIR}$ as shown in
Figure~\ref{fig:sed_stacking_hz}.

To derive realistic uncertainties on star formation rates and
specific star formation rates, we bootstraped galaxies simultaneously
in all the infrared bands, i.e., bootstrapping SEDs. Each time we
randomly selected a subsample of the galaxies, performed median
stacking in all the bands, and determined the SFR and sSFRs from the
TIR and median stellar mass of the subsample. We repeated this
process 50 times and determined the dispersion of SFR and sSFR. This
method avoids the drawback that the uncertainties in different bands
as derived from bootstrapping galaxies in a single band are likely
correlated, e.g., galaxies that are fainter in the shorter wavelength
may tend to be brighter in the larger wavelength due to the
variations in dust temperature and/or redshifts.

\subsection{Dust attenuation of HIEROs}

A tight correlation between
$L_{\rm IR}/L_{\rm UV}$ and UV continuum slope $\beta$ exists for UV-selected L$^{*}$
star-forming galaxies in the local universe and up to $z \sim
2.5$~\citep{Meurer:1999,Kongx:2004,Reddy:2012a}. At higher redshifts,
it becomes more difficult to directly measure both $L_{\rm IR}$ and
$\beta$. Based on infrared stacking analysis, \cite{Lee:2012} derived
IR luminosities for a statistical sample of $L > > L^{*}$,
$z\sim 4$ LBGs and showed that they are consistent with the
IRX--$\beta$ relation presented by \cite{Meurer:1999}.  For HIEROs,
because most of them are relatively faint in the rest-frame UV, we
first performed a median stacking across the observed-frame {\it HST}/ACS
F435W to {\it HST}/WFC3 F160W bands for $JH\text{-}blue$ HIEROs and
$JH\text{-}red$ HIEROs, respectively, and then measured their flux densities
based on the stacked images. We then used the  measurements for
rest-frame 1400--2800~\AA\ to derive 
$\beta$. This process is illustrated in
Figure~\ref{fig:extinction_correction}. The strong break between
observed-frame $B_{435}$-, $V_{606}$-, and $I_{775}$-band for
$JH\text{-}blue$ HIEROs provides additional independent evidence that
most HIEROs should be at similar redshifts as $B$-dropouts and
$V$-dropouts, i.e., $z \sim 4$--5.

The right panel of Figure~\ref{fig:extinction_correction} compares
the relation between $L_{\rm IR}/L_{\rm UV}$ and 
$\beta$ for both HIEROs and LBGs.  The HIEROs have significantly
redder UV slopes ($\beta \sim 0$) than the brightest/most massive
LBGs ($\beta \sim -1.9$) at similar redshifts. On the other hand, the
effective dust attenuation, $L_{IR}/L_{UV}$, is extremely high for
HIEROs, reaching $\sim$1000 for both $JH\text{-}blue$ HIEROs and
$JH\text{-}red$ HIEROs. The $JH\text{-}red$ HIEROs fall on the Meurer
relation within uncertainties while $JH\text{-}blue$ HIEROs tend to
be above the Meurer relation, which is consistent with the results on
a $K_{s}$-selected massive galaxy sample at $z \sim 3.3$ by
\cite{Pannella:2015}. This is at odds with recent findings on
UV-selected samples at $z \sim 5$, which are systematically below the
Meurer relation \citep{Capak:2015}. This may be caused by
UV selection tending to select less massive and less dusty
galaxies, leading to a biased view of galaxy populations at
high redshift.

\subsection{Star formation rates of HIEROs}

\begin{figure*}[htb] 
\centering
\includegraphics[trim=-20 -20 0 0,width=0.49\linewidth]{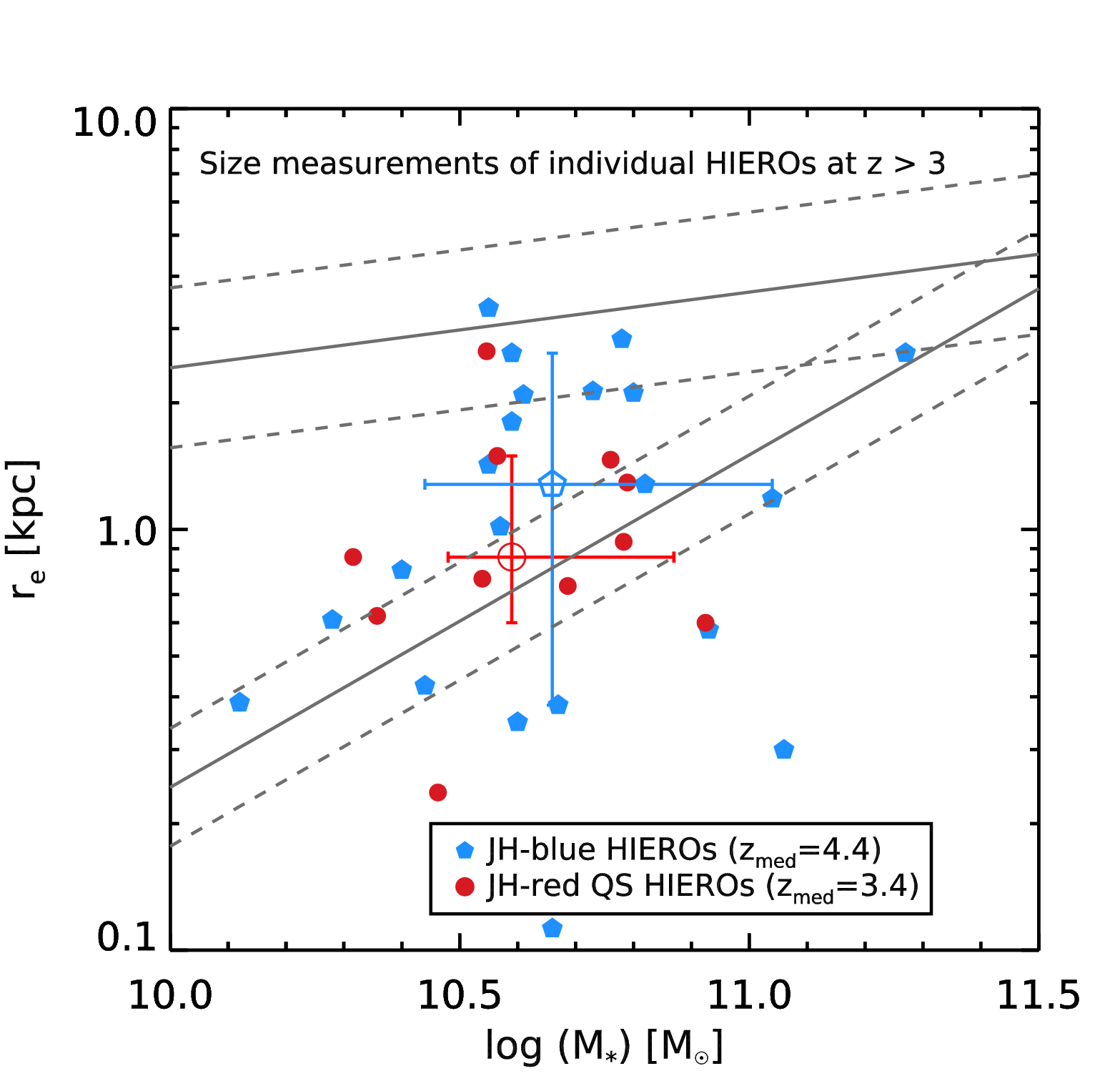}
\includegraphics[trim=-20 -20 0 0,width=0.49\linewidth]{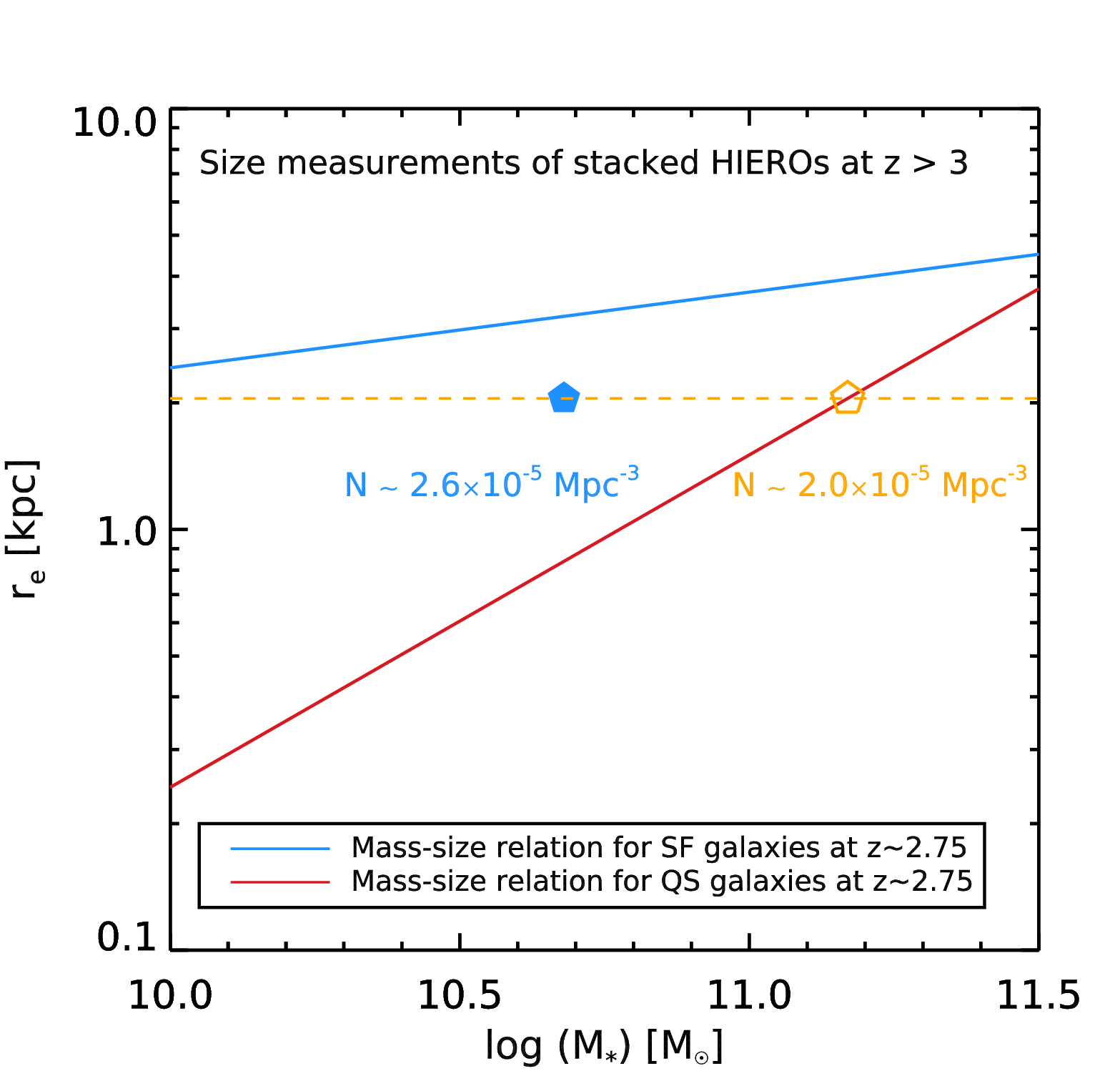}
\caption{\textbf{Left:} Mass-size relation from individual
  measurements of quiescent ($JH\text{-}red$ QS) and
  ($JH\text{-}blue$) star-forming HIEROs at $z > 3$ from the
  {\it HST}/WFC3 F160W band. Here only galaxies with reliable size
  measurements are shown. The open symbols indicate the median size
  and mass and their 16 and 84th percentile for the two
  populations. The grey lines indicate the mass-size relation at $z =
  2.75$ and its associated 1$\sigma$ scatter for quiescent and
  star-forming galaxies \citep{vanderWel:2014}. \textbf{Right:}
  Mass-size relation from stacking of $JH\text{-}blue$ HIEROs ($z
  \sim 4.4$, blue filled pentagons) compared to massive quiescent and
  star-forming galaxies at $z= 2.75$~\citep{vanderWel:2014}. 
  Quiescent galaxies at $z \sim 2.75$ with similar sizes
  as $JH\text{-}blue$ HIEROs have stellar mass $\sim 10^{11.2}$~\Msol\
  and number density of $2.0 \times 10^{-5}$~Mpc$^{-1}$,
  which matches the number density of the $JH\text{-}blue$ HIEROs
  (${\sim}2.6 \times 10^{-5}$~Mpc$^{-3}$).\label{fig:stack_size_HIERO}}
\end{figure*}

Using a \cite{Kennicutt:1998} conversion of SFR [\Msol~yr$^{-1}$] =
$L_{\rm IR}[\Lsol]/10^{10}$ (\cite{Chabrier:2003} IMF), 
the SFR for $JH\text{-}red$ and $JH\text{-}blue$ HIEROs are
120~\Msol~yr$^{-1}$ and 240~\Msol~yr$^{-1}$, respectively. 
These and other properties are listed in
Table~\ref{tab:statistics_sfr}. Figure~\ref{fig:ssfr_z_massive}
shows  the sSFR for $JH\text{-}blue$
and $JH\text{-}red$ SF HIEROs. The
sSFR for $JH\text{-}blue$ HIEROs, 4.2~Gyr$^{-1}$, is twice as high as
that for similarly massive galaxies at $z \sim 2$~\citep{Schreiber:2015}. 
This suggests that the sSFR of massive
galaxies continues to increase at $z > 2.5$ and up to $z \sim
4$. However, the growth rate at $z \gtrsim 2.5$ is slower than that at $z
\sim 0$--2. For instance, \cite{Sargent:2014} found that the evolution
of sSFR at $z \lesssim 2.5$ roughly follows $
(1+z)^{2.8}$, which predicts sSFR $\sim 7$~Gyr$^{-1}$, much higher
than we observed here at $z \sim 4$. Based on simple analytic
arguments on the accretion rates into halos and the accretion of
baryons into galaxies, \cite{Dekel:2014} showed the evolution of sSFR
for typical galaxies should follow $sSFR \sim (1+z)^{2.5}$, which
also predicts a higher sSFR at $z \sim 4$ than observed for HIEROs. Several recent
studies report similar slow evolution of sSFR at $z \gtrsim 2.5$ but
for less massive or for UV-selected
galaxies~\citep[e.g.,][]{Gonzalez:2014,Tasca:2015}. Our finding
suggests that the slow 
evolution of sSFR at $z \gtrsim 2.5$ (or the fast evolution of sSFR
at $z < 2.5$) is likely a universal behavior for all masses of
galaxies.

Most previous studies on the cosmic star formation rate densities at
$z \gtrsim 4$ contain only contributions from LBGs. The cosmic SFRD
contributed by HIEROs needs to be added to have a complete view of
cosmic SFR densities at high redshift.  Using a total area of 
340~arcmin$^{2}$ and assuming that individual HIEROs have the
same sSFR as derived from stacking, the total SFRD contributed by
HIEROs is ${\sim} 6 \times
  10^{-3}$~\Msol~yr$^{-1}$~Mpc$^{-3}$ at $z \sim 3.7$ and 
 ${\sim} 8  \times 10^{-3}$~\Msol~yr$^{-1}$~Mpc$^{-3}$ at $z \sim 4.7$.
By comparison, recent studies showed that the total SFRD of LBGs is
$\sim$0.049 and $\sim$0.036~\Msol~yr$^{-1}$~Mpc$^{-3}$ at
$z\sim 4$ and $z \sim 5$,
respectively \citep{Bouwens:2014}. Thus despite the small
number of HIEROs, they contribute 15--25\% to the SFRD at $z \sim 4$--5,
not taken into account in previous studies based on
UV-selected samples. (There is essentially no overlap between the
HIERO and LBG selections.)  Figure~\ref{fig:ssfr_z_massive} plots the
SFRD from HIEROs identified in this work and from LBGs in previous
work~\citep{Bouwens:2014,Finkelstein:2015}.

\section{Structural properties of HIEROs}
\label{Sec:structure}

Recent work has revealed a significant population of massive compact
quiescent galaxies at $z \sim 1.5$--3 with typical ages of
${\sim} 1$~Gyr \citep{Daddi:2005,Buitrago:2008,Szomoru:2012,Cassata:2013,
vanderWel:2014}.
It is suggested that the formation of these galaxies must have been
through a compact star-forming
phase~\citep{Barro:2013,Patel:2013,Stefanon:2013,Williams:2014,Zolotov:2015}. Indeed,
a significant population of massive compact star-forming galaxies
exists at $z \sim 2$--3 \citep{Barro:2013,Barro:2014}. These are
likely to form the bulk population of compact quiescent galaxies at
$z \lesssim 2$. However, when and how these compact galaxies formed
remains unclear. Moreover, a significant number of passive galaxies
have already formed at $z \sim 3$--4
\citep{Gobat:2012,FanL:2013,Buitrago:2013,Straatman:2014}, as also shown in this
work, and the star-forming progenitors of these earliest quenched
systems remain elusive \citep{Straatman:2015}. Studying structural
properties of $JH\text{-}blue$ HIEROs, which are representative
massive galaxies at $z > 3$, should provide us important insights
into these questions.

For both the $JH\text{-}blue$ and the $JH\text{-}red$ QS HIERO
populations, we limit sample galaxies to those with photometric
redshifts $z > 3$. We then retrieved their F160W-band half-light
radii from the CANDELS structural parameters catalogs as described
by \cite{vanderWel:2014}. The median F160W magnitudes for the
$JH\text{-}blue$ HIERO and $JH\text{-}red$ QS HIERO are $H = 25.5$
mag and $H = 24.5$~mag, respectively. Therefore, in the left panel of
Figure~\ref{fig:stack_size_HIERO} we only plot galaxies that have
reliable size measurements (with flag = 0 in the catalog of
structural parameters, as described in \cite{vanderWel:2012}). This
leaves us 22 (out of 66) $JH\text{-}blue$ and 11 (out of 15)
$JH\text{-}red$ QS HIEROs with respective median F160W magnitudes of
$\langle H\rangle = 25.0$~mag and $\langle H\rangle = 24.5$~mag. Quiescent galaxies at $z \sim 3.4$
are as compact as their $z \sim 2.75$ counterparts.

\begin{figure}[!htb] 
\centering
\includegraphics[trim= 0 40 0 0,clip,width=0.99\linewidth]{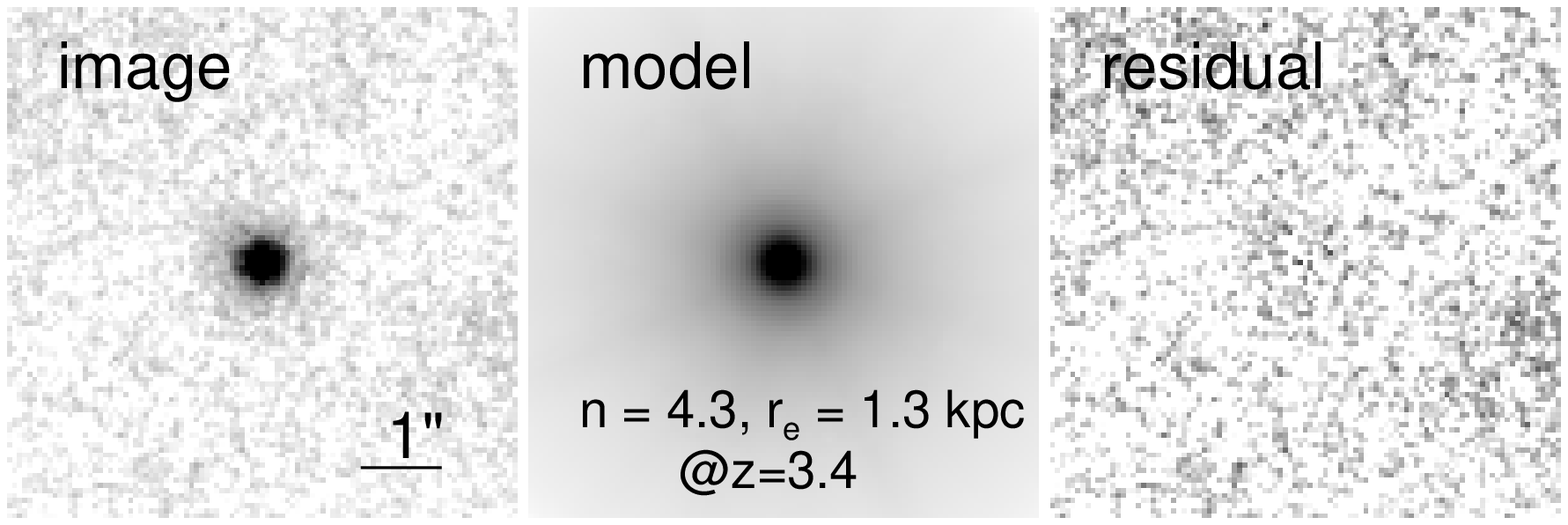}\\
\includegraphics[trim= 0 40 0 0,clip,width=0.99\linewidth]{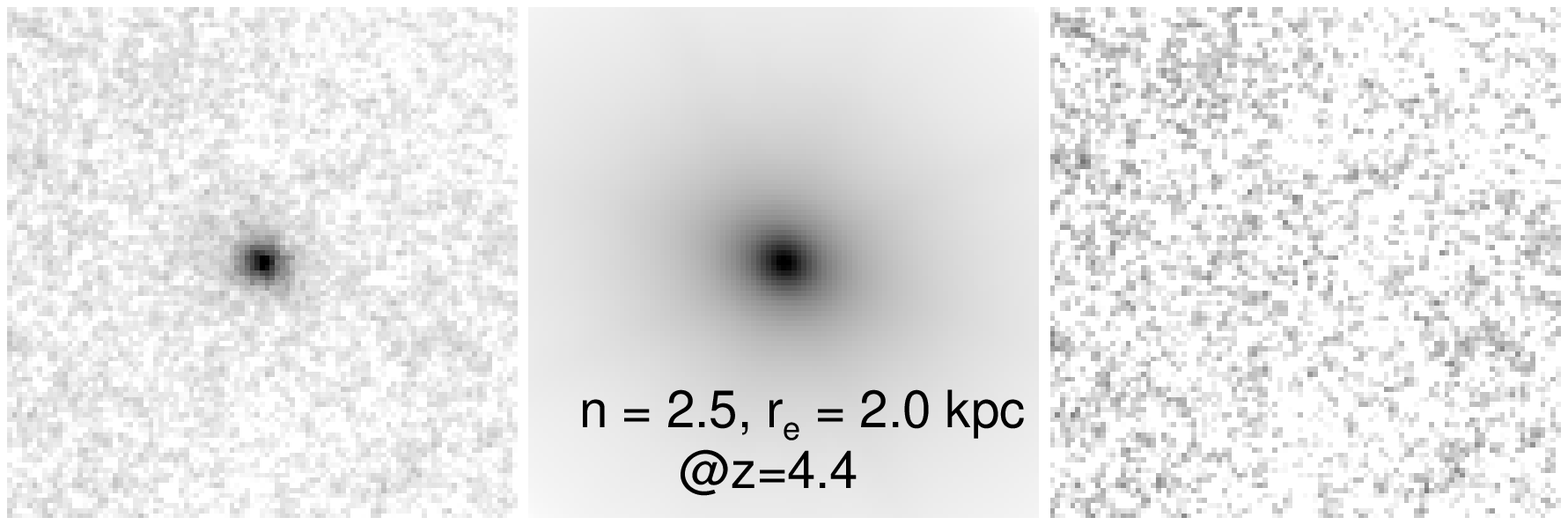}
\caption{S\'{e}rsic fits to the stacked F160W image of passive (top
  panels) and ($JH\text{-}blue$) star-forming (bottom panels)
  HIEROs. Panels from left to right  show the galaxy image, the
  best-fitting model, and the residual (observed minus model) image.
  \label{fig:size_stamp}}
\end{figure}

There are similar numbers of compact star-forming HIEROs compared to
quiescent HIEROs, though the star-forming ones tend to be at higher
redshifts. If we define galaxies below the 1$\sigma$ upper bound of
the mass-size relation for $z \sim 2.75$ quiescent galaxies as
compact, then there are 7/11 compact quiescent and 10/22 compact
star-forming HIEROs, respectively. This corresponds to a number
density of $9.3 \times 10^{-6}$ Mpc$^{-3}$ and $6.7 \times 10^{-6}$
Mpc$^{-3}$ for compact quiescent ($3 < z < 4$) and star-forming
HIEROs ($3.5 < z < 5$), respectively. The actual fraction of compact
galaxies for star-forming HIEROs are likely lower because most of
them are not shown in this figure due to unreliable size
measurements, which tend to have larger sizes based on
stacking.  This seems at odds with the findings
of~\cite{Straatman:2015}, who revealed a paucity of compact
star-forming galaxies in their $z \sim 4$ galaxy samples. We argue
that this is likely due to the difference in the sample selection
methods and cosmic variance. (\citeauthor{Straatman:2015} selected their
sample based on $H$ and $K_{s}$ catalogs and focused on
galaxies with $M_{*} > 10^{10.6}$~\Msol\  at $3.4 < z < 4.2$.)  The
majority of the $JH\text{-}blue$ HIEROs are too faint to have
reliable size measurements, so we measured the size of the stacked
F160W image (Figure~\ref{fig:stack_size_HIERO}). We derived the
half-light radius of the stacked image with single S\'{e}rsic model
fitting using GALFIT \citep{PengC:2010}, as shown in
Figure~\ref{fig:size_stamp}. The best-fit S\'{e}rsic model yields $n
= 2.5$ and $r_{e} = 2.0$~kpc (assuming a median redshift of $z \sim
4.4$) for star-forming HIEROs while it yields $n = 4.3$ and $r_{e} =
1.3$ kpc (assuming a median redshift of $z \sim 3.4$) for quiescent
HIEROs.

Based on the mass-size relation of quiescent galaxies at $z \sim
2.75$ \citep[Figure.~\ref{fig:stack_size_HIERO},][]{vanderWel:2014},
 the stellar mass of quiescent galaxies with similar
sizes is $M_{*} \sim 10^{11.17}$~\Msol. With current SFR of
240~\Msol~yr$^{-1}$, the HIEROs can reach this stellar mass within
$\sim$0.6~Gyr, less than the time interval between the
median redshift of HIEROs ($z \sim 4.4$) and $z \sim 2.75$ quiescent
galaxies.  This time scale, $\sim$ 0.6 Gyr, is similar to the gas
depletion time for typical (main-sequence) star-forming galaxies at
$z \sim 1-3$ \citep{Tacconi:2013}.  We thus conclude that the HIEROs
can evolve into the most massive quiescent galaxies at $z \sim 2.75$
by in situ star formation and subsequent quenching.
Moreover, the number density of massive quiescent galaxies with
$M_{*} > 10^{11.17}$~\Msol\ at $z \sim 2.75$ is ${\sim}2.0 \times
10^{-5}$~Mpc$^{-3}$ \citep{Muzzin:2013b,Ilbert:2013}, similar to the
number density of $JH\text{-}blue$ HIEROs, 2.6 $\times
10^{-5}$~Mpc$^{-3}$. The $JH\text{-}blue$ HIEROs are in general at
much higher redshift than the compact star-forming galaxies
identified by \cite{Barro:2013} and \cite{Williams:2014} and are
also more massive than those identified by \cite{Williams:2014}. We
propose that the HIEROs in our sample likely include the majority of
the progenitors of the most massive, also likely to be the first
quenched, quiescent galaxies at $z \sim 2.75$. Further investigations
of the HIEROs would be key to unveil the formation mechanism of the
earliest-quenched massive (compact) quiescent galaxies.

\section{Completeness of the HIERO selection for
  massive galaxies at $z > 3$} 
\label{Sec:discussion}

The new color selection technique identifies massive red galaxies at
$z > 3$. Both photometric analysis on individual optical-to-NIR SED
and the stacked infrared SED suggest that we reveal a population of
$z > 3$ massive galaxies that was largely missed in previous
UV-selected samples. A critical question is then  to
what extent  these HIEROs represent massive galaxy populations at
$z \gtrsim 3$.

\begin{figure}[!htb] 
\centering
\includegraphics[trim =  0 0 0 0, clip,angle=0,width=0.99\linewidth]{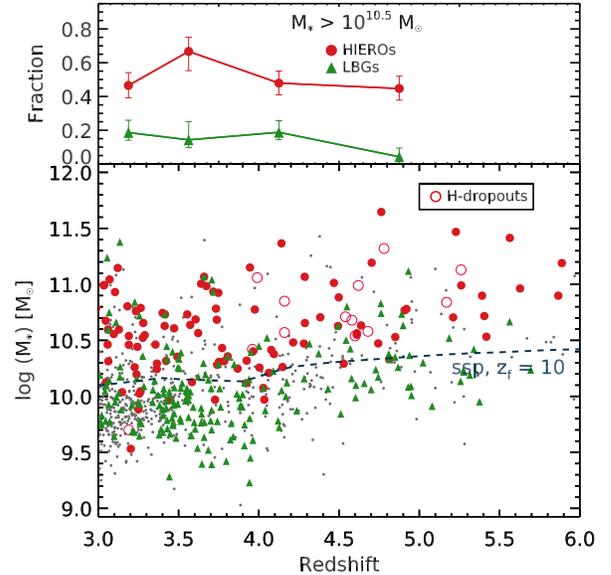}
\caption{\textbf{Bottom panel}: Stellar mass versus redshift for all
  galaxies down to a limiting AB magnitude of $[4.5] = 24$ and $z
  > 3$ in GOODS-South and GOODS-North. The photometric redshifts for
  $H$-detected galaxies are from CANDELS \citep[see, e.g.,
  ][]{Dahlen:2013} while those for $H$-dropouts were derived here with
  FAST\null. HIEROs are denoted in red while LBGs ($B_{435}$-,
  $V_{606}$- or $I_{775}$-dropouts) are denoted in green. The blue
  dashed line shows the mass completeness of our 4.5~$\mu$m-selected sample
  ($[4.5] < 24$) based on an instantaneous-burst BC03 model formed at $z =
  10$. \textbf{Top panel}: The respective fractions of HIEROs and
  LBGs for galaxies with $M_{*} > 10^{10.5}$~\Msol.
\label{Fig:HIERO_pz_mass}}
\end{figure}

\begin{figure}[!htb]
\centering
\includegraphics[width=0.9\linewidth]{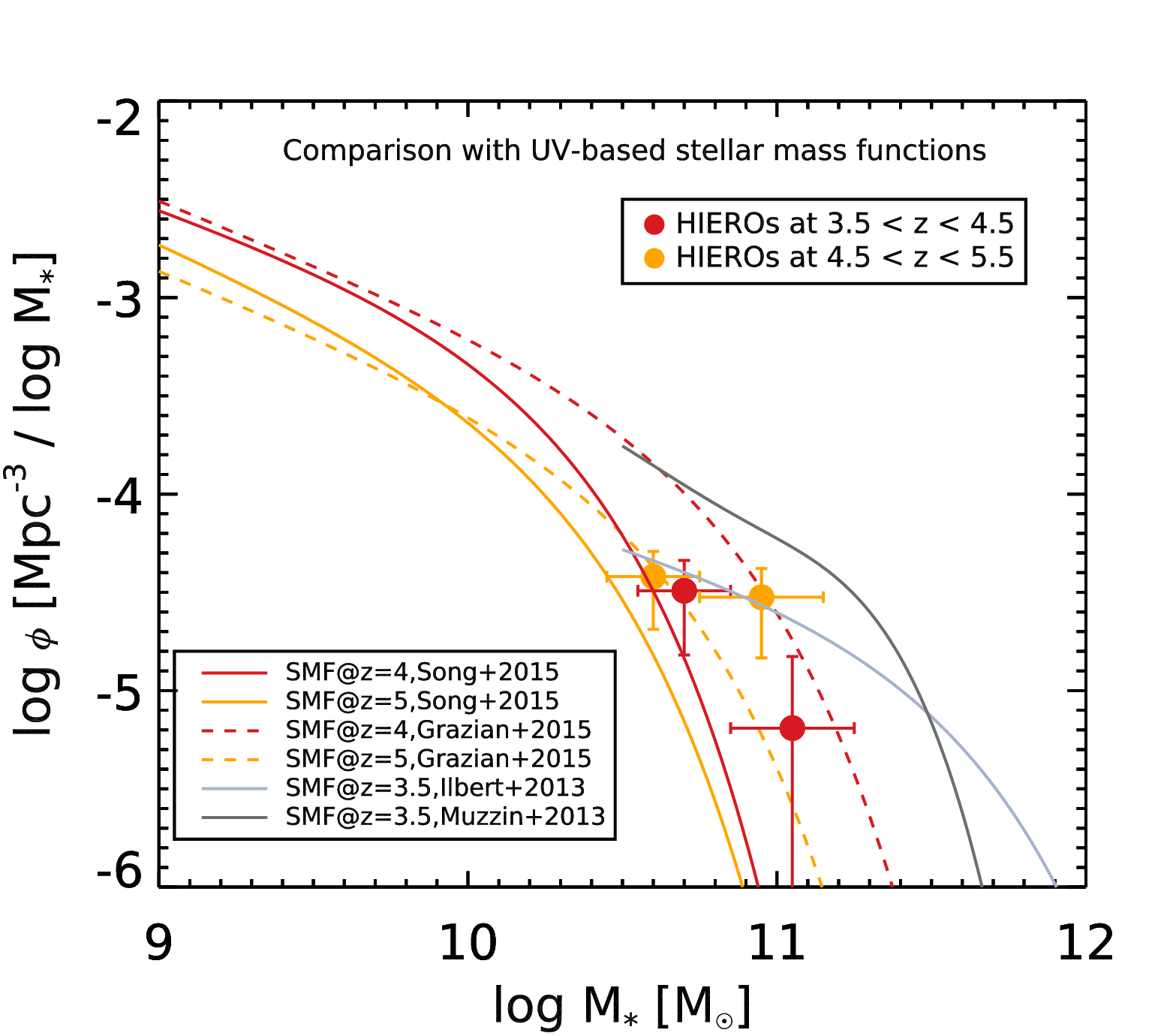}\\
\includegraphics[width=0.9\linewidth]{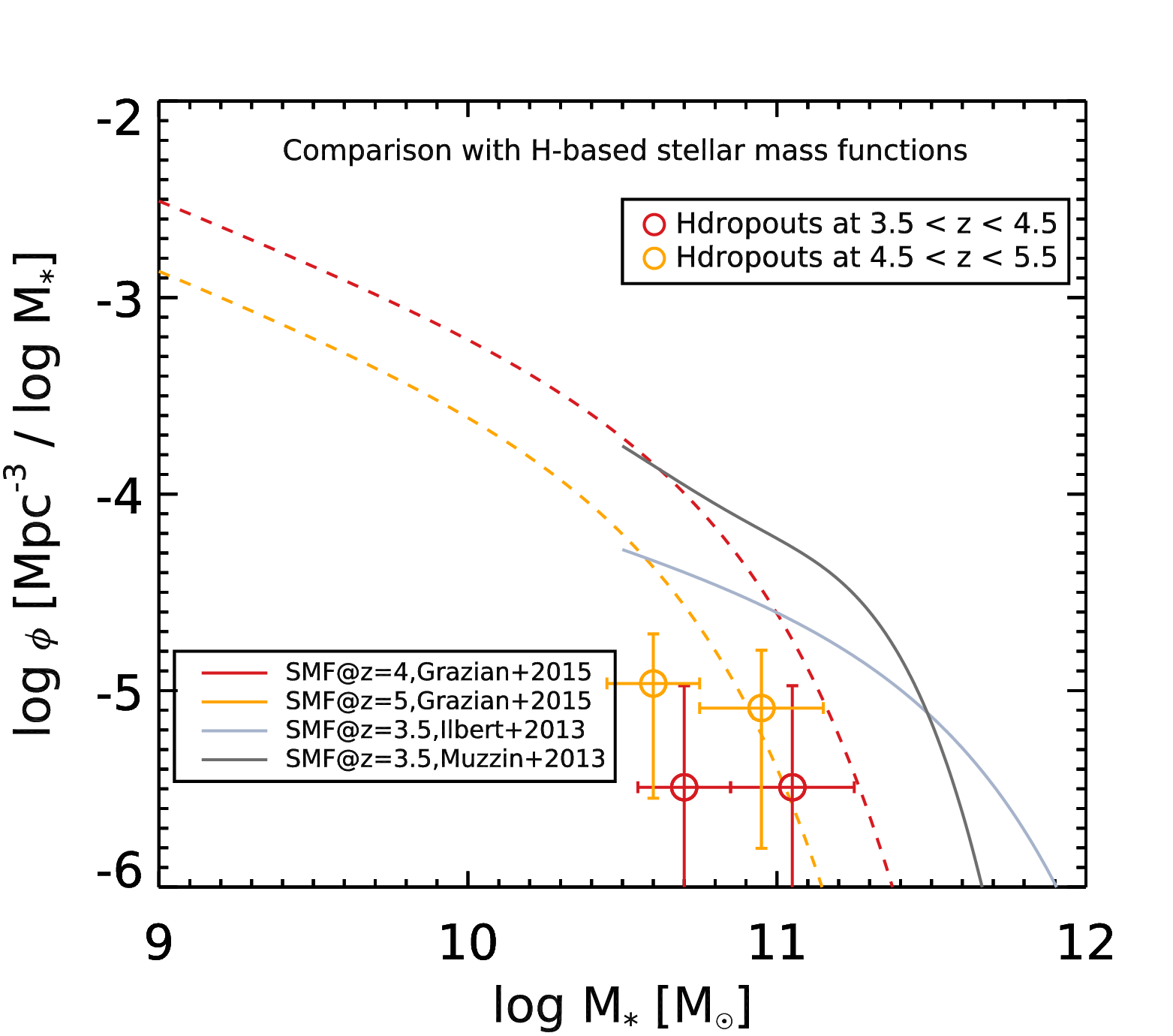}
\caption{Number densities of massive $JH\text{-}blue$ HIEROs (top
  panel) and $H$-dropouts (bottom panel) compared to the stellar mass
  functions based on UV-selected samples \citep{Songm:2015} and
  $H$-band selected samples \citep{Grazian:2015} at $z \sim 4$--5. The
  massive end of the stellar mass function at $3 < z < 4$ from the
  COSMOS survey is also shown \citep{Muzzin:2013b,Ilbert:2013}. The
  HIEROs are separated into two stellar mass bins. 
  \label{fig:mass_functions}}
\end{figure}

Unfortunately it is difficult to precisely estimate the selection
efficiency of HIEROs because there are no complete spectroscopically
confirmed massive galaxy samples at these redshifts. Bearing in mind
the uncertainty in photometric redshifts (especially for those
galaxies that are neither LBGs nor HIEROs), we seek a rough estimate
of the selection efficiency of the HIERO criterion in selecting
massive galaxies at $z > 3$ by comparing to
photometric-redshift-selected samples.
Figure~\ref{Fig:HIERO_pz_mass} presents the stellar mass versus
redshift for all galaxies with $[4.5] < 24$ and $z > 3$ in the two
GOODS fields, which are selected based on the F160W-band selected
catalogs from CANDELS and those $H$-dropouts identified in this
work. For galaxies in the CANDELS catalog, we use the stellar mass
and redshifts from the official CANDELS catalog while for the
$H$-dropouts we use the stellar mass and redshifts estimated in this
work. We emphasize again that no significant differences are found
for $JH\text{-}blue$ HIEROs in terms of their redshift and stellar
mass estimates between this work and those from CANDELS.

Figure~\ref{Fig:HIERO_pz_mass} shows that ${\sim} 60\%$ of the galaxies
with $M > 10^{10.5} $~\Msol\ at $z \gtrsim 3.5$ can be identified
by our HIERO selection, while only ${\sim} 10\%$ are selected as LBGs
($B_{435}$-, $V_{606}$-, or $I_{775}$-dropouts)
using the same criteria as 
\cite{Bouwens:2012}. Moreover, ${\sim} 20\%$ of the massive galaxies
are $H$-dropouts, which are not included in the CANDELS $H$-band-selected
catalog.  To explore whether these results strongly depend on the
photometric redshift estimation, we also derived the fraction of
HIEROs and LBGs at the high mass end in the GOODS-South field using
photometric redshift estimation from \cite{Hsu:2014}, yielding
similar results on the fraction of massive galaxies selected as
$H$-dropouts and HIEROs in general. These results suggest that the
HIEROs dominate the high-mass end of the stellar mass function at $z
> 3$, and their properties are thus representative of the massive
galaxies at these redshifts. On the other hand, the LBG selection
misses the majority of the most massive galaxies at high redshifts
and significantly underestimates the high-mass end of the stellar
mass function. Similarly, using the $H$-band selected samples alone
would also miss a significant fraction of massive galaxies.

Figure~\ref{fig:mass_functions} presents the comparison between the
number density of HIEROs and the galaxy stellar mass functions (SMF)
based on both UV and $H$-selected samples. We separated the HIEROs with
$M_{*} > 10^{10.5}$~\Msol\ into two stellar mass
bins. Figure~\ref{fig:mass_functions} reveals that the HIEROs
dominate at the high-mass end and represent the key population
that reconciles the discrepancies between UV-based
SMFs \citep{Songm:2015} and SMFs based on stellar-mass-selected
samples \citep{Grazian:2015} at $z \sim 4$--5 (but we caution that the uncertainty at the massive end of
the SMF for both estimates is quite large). On the other hand, we
also compare particularly the number density of $H$-dropouts with the
SMF based on $H$-band-selected samples \citep{Grazian:2015}. In this
way we can derive the fraction of massive galaxies
missed by $H$-band selection. As shown in the bottom panel of
Figure~\ref{fig:mass_functions}, at 
$M_{*} > 10^{11}$~\Msol, the number density of $H$-dropouts is
comparable to that of $H$-detected galaxies (down to the limit of the
CANDELS survey). This suggests that, even based on $H$-band-selected
samples (or more precisely, mass-selected samples based on
$H$-band catalogs), we may still substantially underestimate
the most massive end and get a steeper exponential tail while the
true exponential tail may be much shallower. Unfortunately,
spectroscopic confirmation of these most massive galaxies at the
highest redshift is not possible with current optical and NIR
facilities. Instead, JWST and ALMA would be the most promising tools
to fully address this question.

\section{Conclusion}
\label{Sec:conclusion}

This paper demonstrates a new $H$ and IRAC color selection technique
(HIEROs: $H - [4.5] > 2.25$) to identify massive, UV-faint galaxies at
$z > 3$ that are systematically missed by the Lyman-break selection
technique. The HIEROs also include a significant population of
massive and dusty galaxies at $z \sim 2$--3, which can be separated
from the true $z > 3$ star-forming galaxies with an additional $J -
H$ color, enabling a clean selection of (dusty) star-forming galaxies
at $z > 3$.  The HIEROs dominate the high-mass end of the stellar
mass function, making up 60\% of galaxies with $M_{*} > 10^{10.5}$~\Msol\
while LBGs contribute only $10\%$.  The fact that only the
$J$, $H$, and IRAC $4.5~\mu$m bands are involved in this selection
allows us to efficiently select large samples of massive galaxies at
$z > 3$.

The high-redshift nature of HIEROs are independently confirmed
through their stacked UV-to-NIR and FIR SEDs: the stacked rest-frame
UV SED resembles those of $B_{435}$- and $V_{606}$-dropouts while the
stacked FIR SED peaks at 500 $\mu$m. Based on the stacked SEDs, UV
and infrared properties of HIEROs are representative of massive
star-forming galaxies at $z > 3$. They are 2--3 magnitudes fainter in
the rest-frame UV than LBGs with the same stellar mass and tend to be
above the IRX--$\beta$ relation. Thus both stellar mass and star
formation rates based purely on UV are underestimated. The $z > 3$
HIEROs have typical SFR $\sim$240~\Msol~yr$^{-1}$ and sSFR
$\sim$4.2~Gyr$^{-1}$, double the rates for similarly massive $z \sim 2$
galaxies and suggesting that the sSFR for massive galaxies continue to
increase at $z > 2.5$ yet with a decreased growth rate compared to
that at $z < 2.5$. This is consistent with recent findings for less
massive and for UV-selected galaxies.

There are similar numbers of compact quiescent and star-forming
galaxies among the $z > 3$ HIEROs, with the star-forming ones being
at higher redshifts than quiescent galaxies ($z \sim 4.4$ vs.\ $z \sim
3.4$). This suggests that even the earliest quenched systems may have
gone through a compact star-forming phase which started at an even
earlier epoch, providing important constraints on the formation of
compact quiescent galaxies. For the total $z > 3$ HIERO population,
both their number densities and sizes match those of the most massive
($M_{*} > 10^{11.2}$~\Msol) quiescent galaxies at $z \sim 2.75$,
providing the most plausible star-forming progenitors.

The HIERO selection provides a reliable and representative sample of
massive galaxies at $z \gtrsim 3$. Although rough constraints on
their typical physical properties can be obtained through stacking,
their general faintness in the rest-frame UV-to-optical inhibits
accurate determination of the physical properties for individual
galaxies with current optical and NIR facilities. On the other hand,
their brightness in the mid- to far-infrared make JWST and ALMA
promising tools to further explore their nature in great detail.
 
\begin{acknowledgements}

  This work is based on observations taken by the CANDELS Multi-Cycle
  Treasury Program and the 3D-HST Treasury Program with the NASA/ESA
  {\it HST}, which is operated by the Association of Universities for
  Research in Astronomy, Inc., under NASA contract
  NAS5-26555. This work is based in part on observations
    made with the {\it Spitzer Space Telescope}, which is operated by the
    Jet Propulsion Laboratory, California Institute of Technology
    under a contract with NASA. Support for this work was provided by
    NASA through an award issued by JPL/Caltech. 
   The research leading to these results has
    received funding from the European Union Seventh Framework
    Program (FP7/2007-2013) under grant agreement n$^{\circ}$:
    312725 (ASTRODEEP). T.W. acknowledges support for this work from the National
    Natural Science Foundation of China under grants 11303014,
    11133001, and 11273015. P.G. acknowledges support from Spanish
    MINECO grant AYA2012-31277.

Facilities: \facility{HST, Herschel(PACS, SPIRE), Spitzer (IRAC, MIPS)}
\end{acknowledgements}


\end{document}